\documentclass[epj]{svjour}

\usepackage{graphics}
\usepackage{amsmath}
\usepackage{amsmath,amssymb}
\usepackage{textcomp,mathcomp}
\usepackage{times}
\usepackage{textcomp}
\usepackage{xcolor}
\usepackage{color}
\usepackage{version}
\usepackage{graphicx}
\usepackage{hyperref}
\usepackage{amsmath}
\usepackage[utf8]{inputenc}
\usepackage{ar}
\begin{document}
%

\authorrunning{S. Villa \textit{et al.}}
\titlerunning{Nuclear relative stiffness from quantitative microscopy analysis}
\title{Non-invasive measurement of nuclear relative stiffness from quantitative analysis of microscopy data}

\author{Stefano Villa\inst{1}, Andrea Palamidessi\inst{2},
Emanuela Frittoli\inst{2}, Giorgio Scita\inst{2,3}, Roberto Cerbino\inst{4}\thanks{Corresponding author: roberto.cerbino@univie.ac.at}, \and Fabio Giavazzi \inst{1}\thanks{Corresponding author: fabio.giavazzi@unimi.it}
}                     
%
%
\institute{Dipartimento di Biotecnologie Mediche e Medicina Traslazionale, Università degli Studi di Milano, I-20090 Segrate, Italy \and IFOM-FIRC Institute of Molecular Oncology, I-20139 Milano, Italy \and Dipartimento di Oncologia e Emato-Oncologia, Università degli Studi di Milano,  I-20133, Milano, Italy \and University of Vienna, Faculty of Physics, 1090 Vienna, Austria}

\date{Received: date / Revised version: date}
%

\abstract{
The connection between the properties of a cell tissue and those of the single constituent cells remains to be elucidated. At the purely mechanical level, the degree of rigidity of different cellular components, such as the nucleus and the cytoplasm, modulates the interplay between the cell inner processes and the external environment, while simultaneously mediating the mechanical interactions between neighbouring cells. Being able to quantify the correlation between single-cell and tissue properties would improve our mechanobiological understanding of cell tissues. Here we develop a methodology to quantitavely extract a set of structural and motility parameters from the analysis of time-lapse movies of nuclei belonging to jammed and flocking cell monolayers. We then study in detail the correlation between the dynamical state of the tissue and the deformation of the nuclei. We observe that the nuclear deformation rate linearly correlates with the local divergence of the velocity field, which leads to a non-invasive estimate of the \textcolor{black}{elastic} modulus of the nucleus relative to the one of the cytoplasm. We also find that nuclei belonging to flocking monolayers, subjected to larger mechanical perturbations, are about two time stiffer than nuclei belonging to dynamically arrested monolayers, in agreement with atomic force microscopy results. Our results demonstrate a non-invasive route to the determination of nuclear relative stiffness for cells in a monolayer.
}
\maketitle
\section{Introduction}
In all the phases of their life cycle, biological tissues are dynamic entities: during morphogenesis, cells move and find their optimal location and shape in the organized spatial distribution of the tissue; in homeostatic equilibrium, dying cells are continuously extruded and replaced by new ones, but a variety of processes - such as tissue repair, inflammatory response, carcinogenesis, and tumor progression -  can drive the tissue out-of-equilibrium. In all these cases, the dynamic state of the tissue changes as a result of the appropriate physical and chemical stimuli being exchanged at different scales, from the one of the single cells to the one involving the entire tissue. However, the details of this process remaining elusive.

With the eyes of a physicist, these changes of state of the tissue can be thought of as phase transitions that bring the tissue from one state to a different one. Focusing on cancer, it is possible to consider cancer progression towards metastasis and invasion as a sequence of steps involving a transition of the tissue from a solid-like to a fluid-like one; this fluidization transition is also referred to as \textit{unjamming} \cite{park2015unjamming,sadati2013collective,blauth2021jamming,malinverno2017endocytic,palamidessi2019unjamming,kim2020unjamming,oswald2017jamming,ilina2020cell}.
The fluid tissue phenotype is typically accompanied at the single cell level by a softening of the cell, which may impact the infiltration of tumor cells \cite{alibert2017cancer,swaminathan2011mechanical}, and at the same time, by an increased cell motility, which makes it easier for sick cells to invade surrounding healthy tissues. Since forces external to the cells propagate through the cytoplasm directly to the nucleus \cite{wang2009mechanotransduction}, larger stresses within the tissue resulting from unjamming can lead to nuclear envelope rupture and consequently lead to DNA leakage and damage \cite{denais2016nuclear,raab2016escrt,xia2018nuclear,pfeifer2018constricted}. In this framework, nuclear mechanical properties and shape are important regulators of the state of a tissue \cite{grosser2021cell}; for example, non-trivial correlations between nuclear stiffness and tissue dynamics have been recently observed \cite{harada2014nuclear,parreira2021single}.

As a consequence of the complexity of considering cellular tissues as materials, an exhaustive characterization of their mechanical and rheological properties is extremely challenging. One of the main reasons is the intricacy of the cell, which results from the composition of several elements: cell membrane, cytoplasm, cytoskeleton, nuclear envelope, organelles, nucleoplasm, chromatine, \textit{etc.}. Moreover, cell mechanical properties depends on a variety of factors, including the cell status, the substrate, the treatments and the applied stress \cite{fischer2020effect,liu2014situ,fabry2003time,guilak2000viscoelastic,darling2015high,harris2012characterizing}. Modelization of the cell as an ensemble of few elastic and viscous elements is often incomplete and unrealistic \cite{alibert2017cancer}. To further complicate the picture, different techniques for the measurement of the cell viscoelastic properties have been deployed (e.g. AFM \cite{radmacher1992molecules,fischer2020effect,liu2014situ}, micropipette aspiration \cite{mitchison1954mechanical,rowat2005characterization}, optical tweezers \cite{titushkin2006distinct}, magnetic twist cytometry \cite{fabry2003time}), and they generally obtain different values for these properties, as they often probe different cellular components \cite{darling2015high,lim2006mechanical}.

Specifically for the cell nucleus, when considering its stiffness within the force field of a tissue one expects values that differ substantially from the ones obtained experimentally in controlled settings and, moreover, different cell lines can have very different mechanical properties, with cells in similar situations exhibiting nuclei either stiffer or softener than cytoplasm \cite{fischer2020effect}. Recently, Parreira \textit{et al.} \cite{parreira2021single} have shown how the presence of a single stiffer nucleus alters the dynamics of a whole monolayer, impairing collective migration. {Through a refined image segmentation of nuclei, allowing the collection of several dynamical parameters, the Authors show that the decrease in net migration velocity is counterbalanced by an increase of the rotational dynamics of cell nuclei, accompanied by an increment of nuclear deformation. Interestingly, nuclear deformation affects all the cells and not only the first neighbors of the stiffer nucleus, thus suggesting that the presence of the obstacle influence nuclei far away from it.}
These results provide an important motivation for the development of non-invasive \textit{in situ} tools for quantifying the nuclear stiffness as a proxy of mechano-biological alterations of the tissue.

In this work, we propose an automated procedure enabling the extraction of time-resolved information on morphology and dynamics of the nuclei in a monolayer, and the investigation of the interplay between nuclear shape fluctuations and collective motility within the cell aggregate. A custom algorithm is developed to identify, segment and track fluorescent nuclei over time in time-lapse microscopy image sequences. The data are subsequently analyzed to investigate correlations between nuclear deformations and kinematics parameters characterizing the monolayer. 
Our estimators allow extracting non-invasively a robust estimate of the relative stiffness between the nucleus and the cytoplasm. We test our approach with different cell lines approaching cell jamming, as well as on tissues undergoing an unjamming transition \textit{via} flocking induced by the overexpression of the RAB5A protein \cite{malinverno2017endocytic}.

\section{Materials and methods}

\subsection{Sample preparation and imaging \label{sec:sample}}
MCF10A cells were a kind gift of J. S. Brugge (Department of Cell Biology, Harvard Medical School,
Boston, USA) and were maintained in Dulbecco's Modified Eagle Medium: Nutrient Mixture F-12 (DMEM/F12) medium (Gibco) supplemented with $5\%$ horse serum, 0.5 mg ml\textsuperscript{-1} hydrocortisone, 100 ng ml\textsuperscript{-1} cholera toxin, 10 $\mu$g ml\textsuperscript{-1} insulin and 20 ng ml\textsuperscript{-1} EGF. The cell line was authenticated by cell fingerprinting and tested for mycoplasma contamination. Cells were grown at 37 \textdegree C in humidified atmosphere with $5\%$ $\text{CO}_\text{2}$. MCF10A cells were infected with pSLIK-neo-EV (empty vector control) or pSLIK-neo-RAB5A lentiviruses and selected with the appropriate antibiotic to obtain inducible, stable cell lines. Constitutive expression of EGFP-H2B was achieved by retroviral infection of MCF-10A cells with
pBABE-puro-EGFP-H2B vector.

MCF10.DCIS.com cells obtained from Dr John F Marshall (Barts Cancer Institute, London, UK) and were maintained in Dulbecco's Modified Eagle Medium: Nutrient Mixture F-12 (DMEM/F12) medium (Gibco)
supplemented with $5\%$ horse serum, 0.5 mg ml\textsuperscript{-1} hydrocortisone, 10 $\mu$g ml\textsuperscript{-1} insulin and 20 ng ml\textsuperscript{-1} EGF. MCF10.DCIS.com cells were infected with pSLIK-neo-EV (empty vector control) or pSLIK-neo-RAB5A
lentiviruses and selected with the appropriate antibiotic to obtain stable inducible cell lines. Constitutive
expression of mCherry-H2B was achieved by retroviral infection of MCF10DCIS.com cells with pBABE-
puro-mCherry-H2B vector.

MCF10AneoT and MCF10CA1 cells were obtained from Dr. Polin Lisa (Barbara Ann Karmanos Cancer Institute, Detroit, USA)
and were maintained in Dulbecco's Modified Eagle Medium : Nutrient Mixture F-12 (DMEM/F12) medium (Gibco) supplemented with $5\%$ horse serum, 1.05 mM $\text{CaCl}_\text{2}$, 10 mM Hepes, 0.5 mg ml\textsuperscript{-1}
hydrocortisone, 10 $\mu$g ml\textsuperscript{-1} insulin and 20 ng ml\textsuperscript{-1} EGF. Constitutive expression of mCherry-H2B was
achieved by retroviral infection with pBABE-puro-mCherry-H2B vector.

Transfections were performed using either calcium phosphate or FuGENE HD Transfection Reagent (cat. no. E2311, Promega), according to the manufacturer’s instructions. Lentiviral and retroviral infections
were performed as described if Ref. \cite{malinverno2017endocytic}.
Cells were seeded in six-well plates in complete medium and cultured until a uniform monolayer had formed. For all the experiments, the seeding density was $1.5$ × $10^6$ cells per well, with the exception of the experiments indicated as LC (low confluency), for which the seeding density was $0.75$ × $10^5$ cells per well.

RAB5A expression was induced, where indicated, 16 h before performing the experiment by adding fresh complete media supplemented with $2.5$ g ml\textsuperscript{-1} doxycycline to cells. At the time of recording, fresh media containing EGF and doxycycline was added. After cell induction, doxycycline
was maintained in the media for the total duration of the time-lapse experiment. 

An Olympus ScanR
inverted microscope with a 5× objective (experiments on jamming monolayers of MCF10.DCIS.com,
MCF10AneoT, and MCF10CA1a cells) or a 10× objective (all the other
experiments) was used to acquire images with a frame rate of $0.5$ frames/min (experiments on jamming monolayers), $0.1$ frames/min (experiment on MCF10A flocking monolayers) and $0.4$ frames/min
(experiment on MCF10A.DCIS.com flocking monolayers) minutes over a $24$ h period. For each sample, 4 or 5 independent FOVs, much smaller than the entire culture plates and far from the boundaries, are captured. Each FOV is imaged both in fluorescence and in phase-contrast microscopy.
\subsection{Nuclear tracking and segmentation}
\label{tracking}
In order to extract both static and dynamic features of the monolayer at the single nucleus level, a Matlab algorithm was developed for the automatic nuclear segmentation and tracking. The algorithm operates on time-lapse microscopy images of monolayers of cells with fluorescently tagged nuclei. The main steps of the algorithm are detailed below.

\subsubsection{Image background subtraction and registration}

Images are first corrected for background intensity inhomogeneities and microscope stage \textcolor{black}{positioning errors}. 

Background subtraction is obtained through the implementation of the iterative procedure described in Ref. \cite{wahlby2002}. We first perform a cubic interpolation of the intensity in each image, by randomly choosing a set of points within a mask which initially coincides with the whole image. After the interpolation, the mask is redefined by excluding the points  corresponding to high intensity values, which are likely to belong to a nucleus rather than to the background, and a second interpolation is performed based on the redefined mask. After an adequate number of iterations (in our case ten), the mask should include only regions belonging to background, evenly distributed across the image. The final interpolation step provides thus a reliable estimate of the background intensity distribution. This process is repeated over different frames equally spaced in time, covering the whole duration of the experiment.
The average of all the so-obtained background images represents our best estimate for the background intensity distribution, which is thus subtracted to each image in the sequence.

Microscope stage \textcolor{black}{translations between consecutive frames, due to positioning errors while imaging different locations at each time point,} can introduce a bias in the reconstructed monolayer dynamics, by adding a spurious contribution to the velocity within the field of view (FOV). In order to address this issue, the Matlab intensity-based registration function \textit{imregister} is used to recover the displacement $\Delta\Vec{x}_{reg}\left(t_i\right)$ of the FOV center of mass between each pair of consecutive frames $(i-1)$ and $i$, considering pure translations.
We then calculate the trajectory  $\Vec{x}_{reg}\left(t_i\right)=\sum_{i'=2}^{i} \Delta\Vec{x}_{reg}\left(t_{i'}\right)$ of the FOV center of mass.
In the presence of both directed cell migration and stage \textcolor{black}{movement}, $\Vec{x}_{reg}$ can be written as  $\Vec{x}_{reg}\left(t_i\right)=\Vec{x}_{cm}\left(t_i\right)+\Vec{x}_{noise}\left(t_i\right)$, where $\Vec{x}_{noise}\left(t_i\right)$ a the random, delta-correlated, noise associated to stage \textcolor{black}{movement}, and $\Vec{x}_{cm}\left(t_i\right)$ is the genuine displacement of the center of mass of the cell monolayer, which we estimate with a twentieth degree polynomial fit of $\Vec{x}_{reg}$. Subtraction of this term from $\Vec{x}_{reg}$ enables isolating the noise contribution $\Vec{x}_{noise}$.

\begin{figure}
\resizebox{1\columnwidth}{!}{%
  \includegraphics{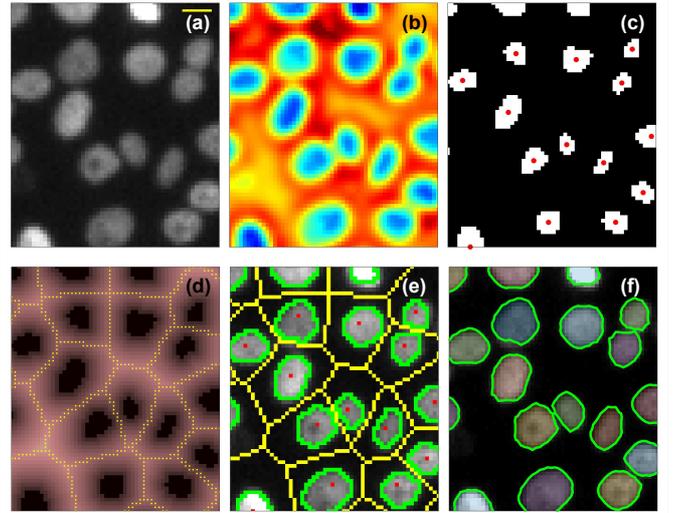}}
\caption{\label{fig_segmsteps}
\textbf{a.} Close-up view of a fluorescence microscopy image of a MCF10.DCIS.com monolayer during a jamming experiment. The yellow bar corresponds to $10$ \textmu{}m.
\textbf{b.} $L_G$ map resulting from the application of LoG filtering to (a), after the Wiener filter application. Colorscale goes from blue to red passing from light-blue and yellow. 
\textbf{c.} Binary map $L_{BW}$ obtained by $L_G$f \text{via} a thresholding operation, as described in the text. Red points marks the center of mass of each connected domain of white pixels.
\textbf{d.} Euclidean distance transform $L_{ED}$ of the $L_{BW}$ map reported in (c). Yellow points mark the result of watershed segmentation of $L_{ED}$.
\textbf{e.} Internal (red) and external (yellow) seed points obtained as shown in panel (c) and (d) respectively are superimposed on the original image. The result of seeded watershed segmentation made on the gradient of (a) is also shown in green.
\textbf{f.} Result of subpixel segmentation at the end of segmentation process.
}
\end{figure}


\subsubsection{Nuclear segmentation procedure}

Once a registered stack of background-subtracted images  $I\left(t_i\right)$ is obtained, we process each frame $i$ to identify single nuclei. Typical nuclei segmentation methods identify nuclei either localizing their centers (for example, looking for intensity maxima or centers of symmetry)  or through the identification of nuclear edges \cite{xing2016robust}. In our case, the identification of the nuclei from their centers is made extremely difficult by the presence of intensity heterogeneity mirroring the chromatin nuclear distribution (Fig. \ref{fig_segmsteps}a); we thus prefer to rely on nuclear edges rather than on nuclear centers. 
To identify nuclear edges, we apply a watershed transform to the image spatial gradient $\nabla I\left(t_i\right)$, which locates nuclear edges of $I\left(t_i\right)$ as ridge lines. However, direct watershed segmentation is not very efficient in properly segmenting the nuclei, in particular in the case of jamming monolayers, where the signal-to-noise ratio is relatively low and partial superposition of two or more nuclei is relatively frequent. In order to improve the quality of the result, we introduce, before the application of watershed transform, a pre-processing where the image  gradient $\nabla I\left(t_i\right)$ is set to zero in correspondence of suitable "seed points" located both within and outside the boundary of each nucleus. The procedure leading to the identification of the seed points is described in detail in the following.

\begin{figure}
\resizebox{1\columnwidth}{!}{%
  \includegraphics{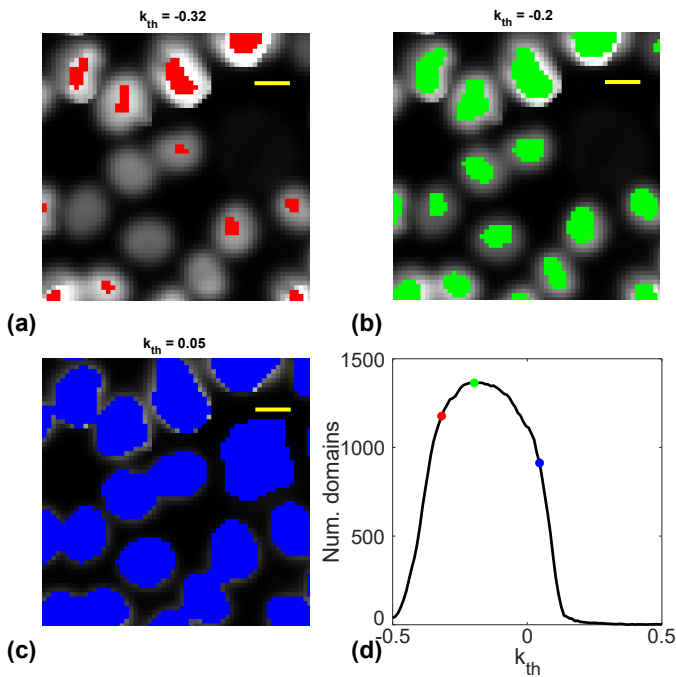}}
\caption{\label{fig_sketch_threshold}
\textbf{a-b-c.} Examples of the details of $L_{BW}$ resulting from the application of different threshold to $L_{G}$ in increasing order of $k_{th}$. Binarized images are overplotted with different colors on the original detail of $I$. Yellow bar correspond to 10 \textmu{}m.
\textbf{d.} Number of connected domains of pixels recovered in $L_{BW}$ as a function of the threshold $k_{th}$. Colored dots mark the same thresholds represented in (a), (b) and (c).
}
\end{figure}

Noise in each frame $I\left(t_i\right)$ is first reduced by applying a Wiener filter, an adaptive noise-removal filtering that preserves nuclei edges \cite{lim1990two}. A Laplacian-of-the-gaussian (LoG) filter is then applied  to enhance nuclear edges. In the obtained map  $L_G$ (see Fig. \ref{fig_segmsteps}.b), higher values (towards yellow), corresponding to nuclear edges of the original image,  surround deep intensity wells (blue), within which the nuclei centers of mass locate. Differences in the fluorescent intensity of different nuclei are reduced by dividing $L_G$ by an intensity map obtained via bicubic interpolation of the local minima of $L_G$. The corrected $L_G$ is then binarized through the application of a suitable threshold $k_{th}$, setting to $0$ ($1$) the pixels whose intensity is larger (lower) than $k_{th}$.   
The threshold value is determined as the one maximizing the number of domains of connected pixels equal to $1$ (Fig. \ref{fig_sketch_threshold}.d). Indeed, as it can be appreciated by comparing figures \ref{fig_sketch_threshold}.a and \ref{fig_sketch_threshold}.b, when, starting from the minimum value of $L_G$, the value of $k_{th}$ is increased, the number of connected domains initially increases as well. This is due to the fact that the correction based on bicubic interpolation of the local minima of $L_G$ does not perfectly level out the intensity differences between nuclei. On the other hand, when $k_{th}$ reaches the typical intra-nuclear value of $L_G$, domains corresponding to different nuclei starts to merge and their number tends to decrease (Fig. \ref{fig_sketch_threshold}.b-c).

Once the proper threshold $k_{th}$ is imposed, the resulting binarized image $L_{BW}$ (Fig. \ref{fig_segmsteps}.c) is an ensemble of domains of connected pixels, each one of which is a first approximation of the inner part of each nucleus.

On $L_{BW}$ additional minor operations are then performed to correct false counting and spurious merging of neighbouring nuclei. False counting are in most cases due to background noise and results in domains of few connected pixels. A morphological "open" operation (an erosion followed by a dilation) has been implemented to remove these artifacts.

\begin{figure}
\resizebox{1\columnwidth}{!}{%
  \includegraphics{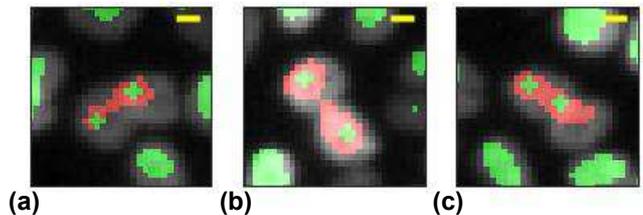}}
\caption{\label{fig_separadoubled}
\textbf{a-c.} Examples of successful separation of domains of connected pixels belonging to different nuclei. Red shadows highlight domains identified with the LoG filter procedure, which are corrected with the subsequent separation step. Green shadows mark the domains determined at the end of the separation procedure. Yellow segments mark 5 \textmu{}m scale.}
\end{figure}

Three examples of mistakenly merged nuclei are marked in red in figure \ref{fig_separadoubled}. In order to correct this problem, the possible cases in which it may be occurred are automatically identified considering the domains with an aspect ratio larger than 3 (against a typical distribution of domains aspect ratios with median around 1.4 and standard deviation of 0.5)\footnote{The domains' aspect ratio limit has been determined through experience, has no real nuclei has been found in the analyzed data set with larger aspect ratios. It can be eventually adjusted for samples with different domains' aspect ratio distribution.}. In order to roughly localize in them the centers of the merged nuclei, for each selected domain the two deeper local minima of $L_G$ are identified and two disks of radius 2 pixels are centered on them, replacing the connected pixels domain. The result of such correction is illustrated in figure \ref{fig_separadoubled}, where bi-nuclear domains are outlined in red, while the domains obtained upon correction are marked in green\footnote{Note that for a proper segmentation the non-perfect centering of the domain within the nucleus does not represent a problem: it is enough that each nucleus contains one and only one entire domain. The precise center of each nucleus is determined after nuclear segmentation.}.

The geometrical centers of mass of the domains of connected pixels are chosen as inner seeds for the watershed segmentation (red points in Fig.\ref{fig_segmsteps}.c).

The determination of the seeds external to the nuclei is simpler. The connected domains in $L_{BW}$ do not only indicate the position of the corresponding nuclei, but have also an area which is typically larger for nuclei with wider projected area. Consequently, the space between two nuclei in $I\left(t_i\right)$ is typically equidistant from the edges of the corresponding connected domains in $L_{BW}$. For the identification of the external seeds, we first determine the Euclidean distance transform $L_{ED}$ of $L_{BW}$ \cite{maurer2003linear}, a map where each pixel value is the distance from the closest domain (Fig.\ref{fig_segmsteps}.d). On $L_{ED}$ a watershed segmentation is finally applied to determine the equidistance lines between noighbouring domains in $L_{BW}$ (yellow points in Fig.\ref{fig_segmsteps}.d), which are then used as external seeds. 

In figure \ref{fig_segmsteps}.e the recovered internal (red) and external (yellow) seeds are superimposed on the original image $I\left(t_i\right)$. By setting  $\nabla I\left(t_i\right)$ to zero in correspondence of all seeds and evaluating the watershed transform, the nuclei segmentation is finally obtained (Fig. \ref{fig_segmsteps}.d, green pixels). The recovered segmentation is pixel resolved; sub-pixel resolution can be achieved by running the Matlab code \textit{subpixelEdges 2.13} \cite{trujillo2013accurate} on the obtained profiles (Fig. \ref{fig_segmsteps}.e).



After nuclei segmentation, several static quantities are easily accessible. In this work, we consider in particular the center of mass $\Vec{x}_j$, the projected area $A_j$ and  the aspect ratio $\AR_j$ of the $j$-th nucleus. The latter is  defined as the ratio between the nuclear major and minor axis, evaluated as the square root of the eigenvalues of the covariance matrix of the segmented objects \cite{jolliffe1986principal}.

\subsubsection{Dynamic parameters}
\label{sec_dynpar}
Time-resolved information on nuclear shape and mobility is recovered by linking the positions $\Vec{x}_j$ of each nucleus in subsequent frames. To this aim we employed the publicly available Matlab implementation by D Blair and E Dufresne \cite{blair2008matlab} of the linking algorithm of JC Grier and DG Crocker \cite{crocker1996methods}. 
Once nuclear trajectories are built, the time evolution of the single-nucleus parameters $\Vec{x}_{j}\left(t_i\right)$, $A_{j}\left(t_i\right)$, $p_{j}\left(t_i\right)$ and $\AR_{j}\left(t_i\right)$ can be also obtained. The instantaneous velocity on the $j$-th nucleus across frames $i$ and $(i+1)$ is estimated as 
$\Vec{v}_{j}\left(t\right)=\left[\Vec{x}_{j}\left(t_{i+1}\right)-\Vec{x}_{j}\left(t_{i}\right)\right]/\delta t$,
where $\delta t$ is the time between two acquired frames. 

\textcolor{black}{Our definition of the instantaneous velocity in terms of the average velocity over a time interval equal to $\delta t$ relies on the fact that, on this timescale, the cellular motion can be assumed to take place with approximately constant velocity. We checked this assumption by measuring the nuclear mean square displacement and estimating the characteristic persistence time of the ballistic-like motion observed for short time delays. As discussed in detail in Appendix \ref{appe_MSD}, for all the considered data sets, the persistence time was found to be larger than $\delta t$.}

The velocity of the center of mass of the monolayer is evaluated as the instantaneous mean velocity of the cells within the FOV 
\begin{equation}
\Vec{v}_{CM}\left(t\right) = \left<\Vec{v}_{j}\left(t\right)\right>_j,
\end{equation}
where $\left<...\right>_j$ denotes the average over all the nuclei in the field of view. The amplitude of velocity fluctuations in the monolayer is evaluated as the root mean square velocity of the nuclei in the center of mass reference frame: 

\begin{equation}
v_{RMS}\left(t\right) = \sqrt{\left<|\Vec{v}_j\left(t\right)-\Vec{v}_{CM}\left(t\right)|^2\right>_j}.
\end{equation}

Comparison between nuclei tracked automatically by the algorithm and manually by an operator reveals that the described segmentation procedure is effective in identifying about $80$-$90$\% and $90$-$95$\% of the nuclei present in each FOV for the jamming and the flocking monolayers, respectively. This difference in tracking efficiency reflects differences in noise level and spatial resolution between different experiments (see previous section for details). Despite the effort to minimize segmentation artifacts, such as multiple segmentation of the same nucleus and spurious merging of multiple nuclei, such segmentation errors occur, especially in those cases where the signal-to-noise ratio is low or partial superposition of different nuclei is frequent. To minimize the impact of segmentation errors on the analysis of nuclear features, a "quality check" has been implemented to select a subset of reliably segmented nuclei. To this end, we evaluate the total instantaneous intensity $J_j\left(t_i\right)$, which is obtained as the sum of the intensity of all pixel within the segmented area of the $j$-th nucleus at frame $i$. The instantaneous value $J_j\left(t_i\right)$ is then compared with its median evaluated over the previous 10 frames. If the difference between $J_j\left(t_i\right)$ and the median is larger than $10$\%, the segmentation of the $j$-th nucleus at frame $i$ is considered unreliable, and the corresponding parameters are not included in the statistics. Trajectories which, after the application of this quality filter, lose more than $20$\% of frames are entirely excluded. 
The number of nuclei that "pass" the quality check varies from sample to sample {between 600 and 5000, depending on the FOVs size and image quality}.

\subsection{Particle image velocimetry}
Particle image velocimetry (PIV) of fluorescent microscopy images of confluent monolayers is performed by using the Matlab \textit{PIVLab} software \cite{thielicke2021particle}.
We choose an interrogation area with size slightly larger than the average inter-nuclear distance, corresponding to approximately $14$ \textmu{}m. Outliers in the reconstructed velocity field, whose components $v_x$ and $v_y$ exceed fixed threshold values, are identified and replaced with the median value of the velocity over neighboring grid points \footnote{For each component $v_{\xi}$ of the velocity, upper and lower bounds are identified as $med(v_\xi) - 10\cdot\sigma_{v_\xi}$ and $med(v_\xi) + 10\cdot\sigma_{v_\xi}$, respectively, where $med(v_\xi)$ and $\sigma_{v_\xi}$ are the median and the standard deviation of $v_{\xi}$ evaluated over the entire time lapse .}. 
From the velocity field obtained from PIV, $\Vec{v}_{CM}\left(t\right)$ and $v_{RMS}\left(t\right)$ are recovered as for PT averaging over grid points coordinates $\Vec{x}$ in place of $j$. Additionally, the local divergence of the velocity field $\nabla\cdot\Vec{v}\left(\Vec{x},t\right)$ is computed from $v_x\left(\Vec{x},t\right)$ and $v_y\left(\Vec{x},t\right)$ using the Matlab \textit{divergence} function.

For the same samples, the PIV algorithm has been tested on both fluorescent and phase contrast time lapses.
As shown in appendix \ref{appe_PIV}, both the PIV analysis return velocity mean values and distributions in agreement with the ones obtained from PT, thus validating the obtained dynamics.


\section{Results and discussion}

\subsection{Jamming monolayers}

A first set of experiments was performed on mature highly confluent monolayers undergoing a jamming transition, a progressive slowing down towards a dynamically arrested state \cite{angelini2011glass,garcia2015physics,cerbino2021disentangling}.

We considered three distinct MCF10A-derived cell lines (MCF10.DCIS.com, MCF10AneoT, and MCF10CA1a cells), each one of those seeded at two different densities (see section \ref{sec:sample} for details).

\subsubsection{Confluent monolayers evolve toward a high density kinetically arrested state}

\begin{figure}
    \centering
    \includegraphics[width=\columnwidth]{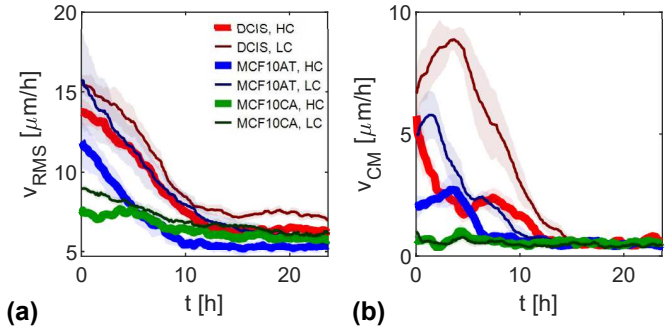}
    \caption{Time evolution of RMS velocity ${v}_{RMS}$ (\textbf{a}) and of the modulus of the center of mass velocity (\textbf{b}) obtained from PIV on fluorescent time lapses. Shadowed error bars are evaluated as the standard deviation of the mean evaluated for each sample over the 5 different FOVs. Reported data refer to jamming samples: DCIS HC (red thick line) and LC (dark red thin line), MCF10AT HC (blue thick line) and LC (dark blue thin line), MCF10CA HC (green thick line) and LC (dark green thin line).}
    \label{fig_vrms_vcm_PIV}
\end{figure}

In figure \ref{fig_vrms_vcm_PIV} we report the time evolution of the RMS velocity ${v}_{RMS}$ (fig. \ref{fig_vrms_vcm_PIV}.a) and of the center of mass velocity (fig. \ref{fig_vrms_vcm_PIV}.b) for the jamming monolayers, as obtained from PIV. Each color refers to a different cell line (thick and thin lines correspond to  high and low seeding density, respectively). For each cell line and seeding density, 5 independent FOVs are considered. 

The general decreasing trend of both the considered indicators towards a plateau follows the expected dynamic arrest characterizing jamming transition and is accompanied by a progressive reduction of the mean cell area, evaluated of the order of $30\%$ \footnote{The cell area reduction as been evaluated through a manual counting of the number of nuclei within a FOV of fixed size at the beginning and at the end of the time lapses.}, as the number of cells grow.

The expected increasing density within the cell monolayer can be observed in figure \ref{fig_A_si}.a-d where four different snapshots of the same FOV of an MCF10A.DCIS.com monolayer are reported at times $0$ h, $8$ h, $16$ h and $24$ h.

\subsubsection{Dynamic arrest correlates with a nuclear projected area decrease and nuclear aspect ratio increase}

\begin{figure}
\resizebox{1\columnwidth}{!}{%
  \includegraphics{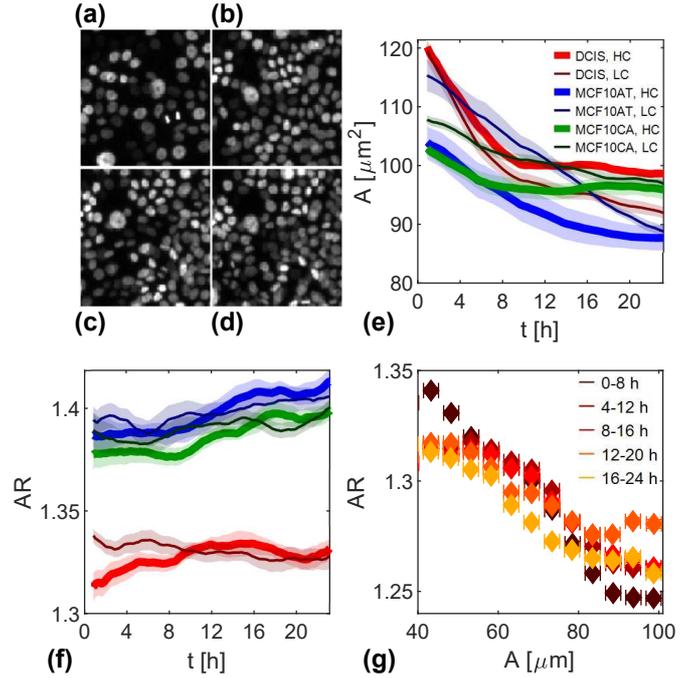}}
\caption{\label{fig_A_si}
\textbf{a-d.} Four different snapshot of the same FOV region of DCIS HC jamming experiment. Snapshots are taken at times $0$ h (a), $8$ h (b), $16$ h (c) and $24$ h (d).
\textbf{e, f.} Time evolution of the projected nuclear area (e) and aspect ratio (f) for DCIS HC (red thick line) and LC (dark red thin line), for MCF10AT HC (blue thick line) and LC (dark blue thin line) and for MCF10CA HC (green thick line) and LC (dark green thin line).
\textbf{g.} Scatter plot of single cell instantaneous measured projected area versus the corresponding aspect ratio. Data are obtained from a DCIS HC jamming monolayer and colors, in order from dark red to yellow, refer to time intervals $0$-$8$h, $4$-$12$h, $8$-$16$h, $12$-$20$h and $16$-$24$h, respectively. Reported points are averages evaluated over equally spaced bins along the horizontal axis.
}
\end{figure}

Nuclear average projected area, reported in figure \ref{fig_A_si}.e, has the same decreasing trend in time of cell area, progressively reducing due to cell proliferation.


Besides these trends, we are here interested in understanding the interplay between the dynamics of the monolayer and changes in nuclear shape at the single cell level.

\begin{figure}
  \includegraphics[width=0.5\textwidth]{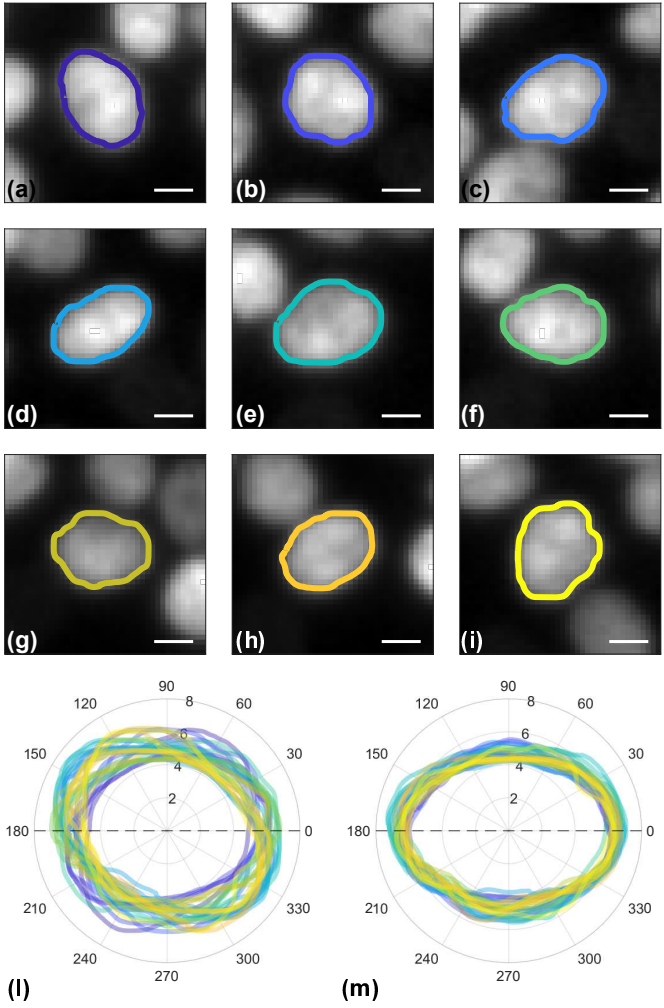}
\caption{\label{fig_segmoverl}
Example of resulting subpixel segmentation of a nucleus followed in time. Images obtained from MCF10A RAB5A overexpressing monolayer in time interval $0$-$5$ hours. Color scale indicates elapsing time from blue to yellow.
\textbf{a-i.} Different images centered on the same tracked nucleus followed in time (timestep $30$ minutes) with the corresponding segmentation overplotted on the original image. White scale bar correspond to $5$ \textmu{}m.
\textbf{l.} Overplot of segmentations at different times (time step $10$ minutes) relative to the same nucleus of (a)-(i). Radial axis units are micrometers.
\textbf{m.} The same of (l), but here each segmentation is corrected by a rigid rotation equal to the recovered nuclear orientation in order to remove rotational contribution and only visualize deformations.}
\end{figure}

On top of the monotonic trends described above, each nucleus undergoes shape fluctuations, as it can be seen in figure \ref{fig_segmoverl} where the contours of the same nucleus are reported for different times. 

Two parameters we have direct access to in order to characterize nuclear shape fluctuations are the area and the aspect ratio. They appear to be strongly correlated. As it can be appreciated from figure \ref{fig_A_si}.f, where we report the time evolution of the average nuclear aspect ratio, the aspect ratio displays a monotonically increasing trend with time. This indicates that, while nuclear projected area reduces as the density increases, nuclear shape anisotropy tends to increase. In order to deepen this observation, we consider the association between the instantaneous aspect ratio $\AR_j$ and the instantaneous area $A_j$ at the single nucleus level. An example of the  resulting scatter plot, binned in equally spaced intervals of area, is reported in figure \ref{fig_A_si}.g. Different colors refer to 5 subsequent and overlapping time intervals of $8$ hours covering the whole duration (24h) of the experiment. A striking negative correlation is observed, proving that in nuclei smaller projected areas are systematically more anisotropic. This observation is in agreement with what observed in Ref. \cite{haase2016extracellular}.
No great differences emerge for different times, suggesting that, at a first glance, the correlation between $A$ and $\AR$ is independent on age and monolayer dynamics.


Since the two parameters are strongly correlated, in the following we only focus on the projected area, whose value can be determined with lower uncertainty.

\subsubsection{Nuclear deformation rates}

Due to the non stationarity of the mean projected area on the experimental time scales, deformation amplitudes are difficult to be uniquely retrieved, as the obtained value are strongly dependent on the way the mean dynamics of the projected area is subtracted.
In analogy with the mean squared displacement, we therefore introduce the mean square strain as

\begin{equation}
    MSS\left(\tau\right) = \left<\left<\Delta a_j^2\left(\tau | t\right)\right>_t\right>_j,    
\end{equation}

where the nuclear strain $\Delta a_j\left(\tau | t\right)$ of the $j$-th nucleus between time $t$ and $t+\tau$ is calculated as

\begin{equation}
    \Delta a_j\left(\tau | t\right) = \frac{\left[A_j\left(t+\tau\right) - A_j\left(t\right)\right]}{\left<A_j\left(t\right)\right>_t}. 
\end{equation}

\begin{figure*}
\includegraphics[width=\textwidth]{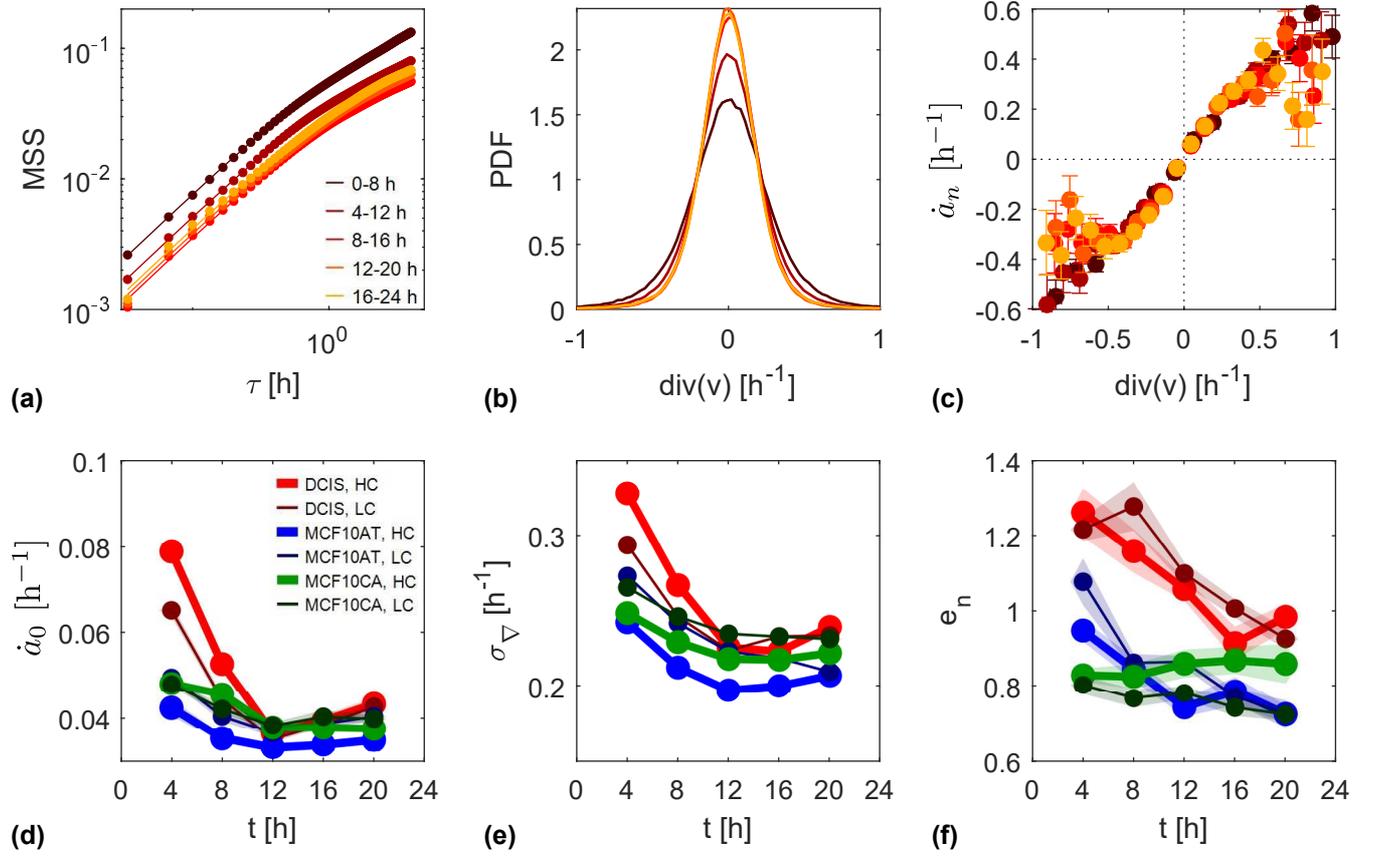}
\caption{\label{Jam_DCIS_HC}
\textbf{a.} Mean square normalized area relative to jamming DCIS HC monolayers. Different colors, in order from dark red to yellow, refer to time intervals $0$-$8$h, $4$-$12$h, $8$-$16$h, $12$-$20$h and $16$-$24$h, respectively. Reported data are scaled to the offset noise $\sigma_w$ obtained from the fit. Lines represent corresponding fits with the model function described in the text, also scaled on $\sigma_w$.  
\textbf{b.} Distribution of the divergence evaluated interpolating on nuclei centers of mass data from PIV on fluorescent images. Data refer to the same sample and the same time intervals of (a).
\textbf{c.} Scatter plot of the instantaneous single nucleus strain rate $\dot{a}_n$ as a function of the corresponding divergence. Reported points are average over hundreds/thousands of strain rate-divergence couples made over bins equally spaced along the abscissae axis. Error bars are evaluated as standard deviations of the mean of the data contained in the bins. Data refer to the same sample and the same time intervals of (a).
\textbf{d.} Nuclear characteristic strain rates as a function of time, obtained from the fit of the corresponding MSS relative to the jamming experiments. Time intervals are the same of (a).
\textbf{e.} Divergence standard deviation as a function of time. Time intervals are the same of (a). 
\textbf{f.} Reciprocal $e_n$ of the slope of $\dot{a}_n$ versus $\nabla\cdot\Vec{v}$ scatter plot. Time intervals are the same of (a).}
\end{figure*}

Representative $MSS\left(\tau\right)$ curves obtained for the DCIS HC monolayer are reported in figure \ref{Jam_DCIS_HC}.a. Each curve refers to one of the five, partially overlapping time intervals the experimental window was divided into. The linearity of $MSS\left(\tau\right)$ at low $\tau$ points to a diffusive-like evolution of the area at short time scales. For larger time delays, $MSS\left(\tau\right)$ becomes sublinear. The limited experimental time window does not allow establishing whether an asymptotic plateau value $\sim1$ is eventually attained. 
In order to extract the key parameters characterizing nuclear deformation, we use an exponential model function $MSS\left(\tau\right) = \sigma_w + \dot{a}_0 \tau_c \left(1-\-e^{-\tau/\tau_c}\right)$ to fit the data. The diffusive-like growth of the area fluctuations is captured by the model in the limit of small $\tau$: $MSS\left(\tau\right) \sim \dot{a}_0 \tau$, with a characteristic strain rate $\dot{a}_0$. The short-time regime is followed by an exponential-like relaxation trend towards a plateau value $\dot{a}_0 \tau_c$. 
The model also includes an offset $\sigma_w$ accounting for random, delta-correlated noise associated to the determination of projected area.

In figure \ref{Jam_DCIS_HC}.d we report the values of $\dot{a}_0$ obtained from the fitting procedure for all the jamming experiments, at different time points. A common feature that can be observed in all datasets, is the progressive decrease in the nuclear strain rate over time, indicating that mechanical deformation of nuclei decreases as kinetic arrest is approached (see also figure \ref{fig_vrms_vcm_PIV}).

\subsubsection{Nuclear deformation rates correlate with the local dynamics of the monolayer}
\label{section_corrVelNuc}

\begin{figure}
\resizebox{1\columnwidth}{!}{%
  \includegraphics{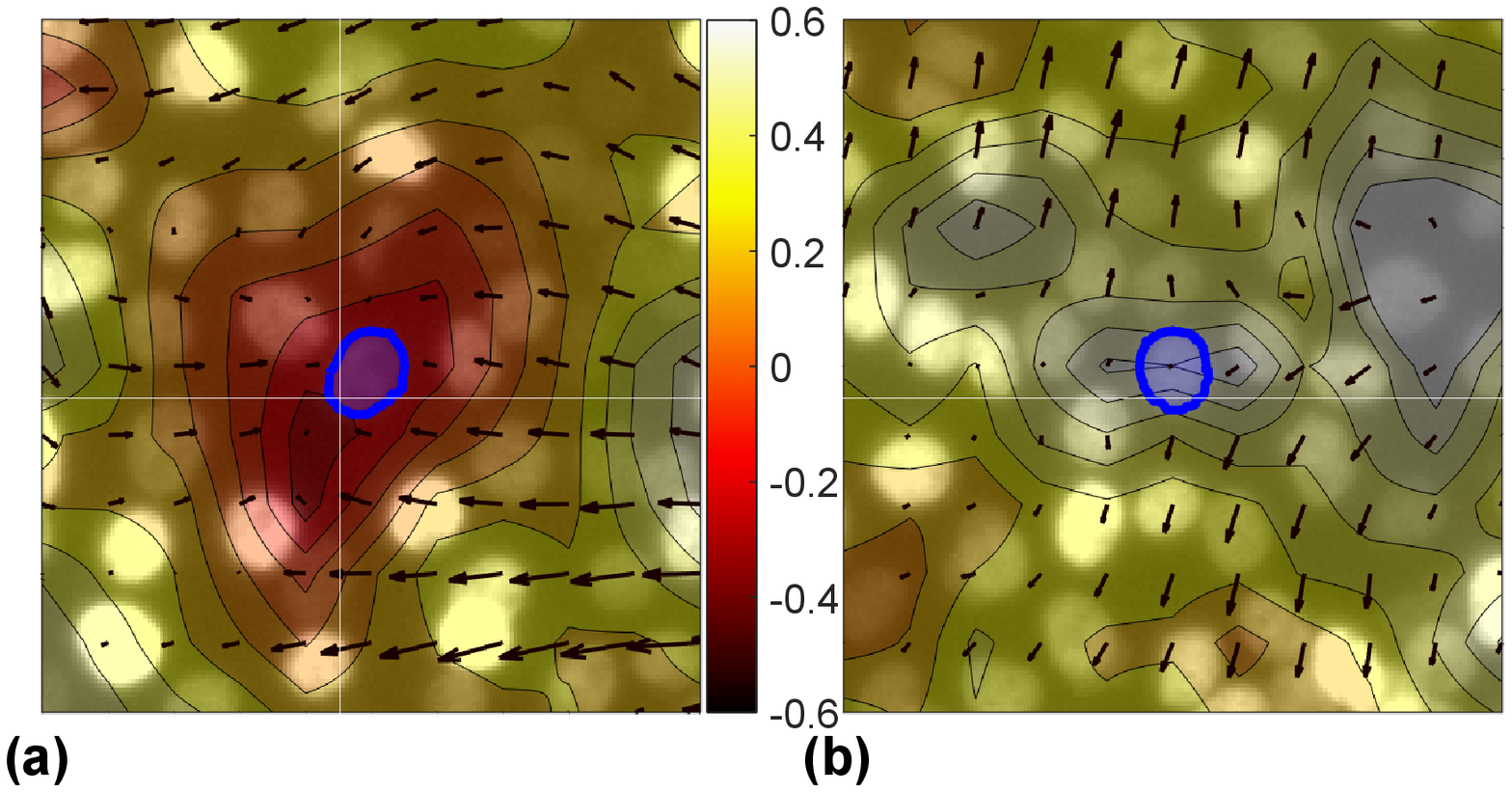}}
\caption{\label{fig_divcompr}
\textbf{a,b.} Overplot of the divergence field over two FOV of an MCF10A cell monolayer. Divergence filed is represented as a shaded colormap where the passage from negative to positive values of divergence are marked by the color progression from red to white. Arrows represent the velocity field. Blue shades higlights two segmented nuclei subjected to a negative (a) and to a positive (b) divergence.}
\end{figure}

In order to investigate in more detail the interplay between nuclear deformation and cell motility, we consider the divergence of the velocity field as a suitable parameter to capture local density fluctuations (\textit{i.e.} compressions or dilations) within the monolayer, as schematically illustrated in figure \ref{fig_divcompr}. We evaluate the divergence field by computing for each time $t$ the divergence $\nabla\cdot\Vec{v}\left(t,\Vec{x}\right)$ of the velocity field obtained from PIV analysis (see appendix  \ref{appe_PIV} for details). We then estimated the value $\nabla\cdot\Vec{v}_j\left(t\right)$ of the divergence in the position $\Vec{x}_j\left(t\right)$ corresponding to the center of mass of the tracked nuclei \textit{via} cubic interpolation.

In figure \ref{Jam_DCIS_HC}.b we report the experimentally determined probability distribution functions (PDFs) of  $\nabla\cdot\Vec{v}_j\left(t\right)$ for the DCIS HC monolayer over the same time intervals considered also in panel a. We observe that the PDF of the divergence tends to become narrower and narrower as the monolayer ages. This can be also appreciated from figure \ref{Jam_DCIS_HC}.e, where the standard deviation $\sigma_{\nabla}$ of the PDF of the divergence is reported as a function of time. The overall trend is very similar to the one found for $\dot{a}_0$.

The origin of the similarity  between the average values reported in figures \ref{Jam_DCIS_HC}.d and \ref{Jam_DCIS_HC}.e can be investigated by considering the direct association between the projected area of the $j$-th nucleus at a given time and the corresponding instantaneous local value of the divergence $\nabla\cdot\Vec{v}_j\left(t\right)$. To this end, we computed the instantaneous single nucleus area strain rate, defined as

\begin{equation}
\dot{a}_j\left(t\right) = \frac{2}{\textcolor{black}{\delta t}}\frac{A_j\left(t_{i+1}\right) - A_j\left(t_{i}\right)}{A_j\left(t_{i+1}\right) + A_j\left(t_{i}\right)}
\end{equation}

In order to probe the correlation between $\dot{a}_j\left(t\right)$ and $\nabla\cdot\Vec{v}_j\left(t\right)$, we evaluate the corresponding time cross-correlation function $K_{\dot{a},\nabla\Vec{v}}\left(\tau\right)$ averaged over the nuclei in the FOVs. For the precise definition of $K_{\dot{a},\nabla\Vec{v}}\left(\tau\right)$ and for consideration and comparison with corresponding self correlations, see appendix \ref{appe_corr}.
In figure \ref{fig_divdacorr}.c of appendix \ref{appe_corr} we report the cross-correlation function for both negative and positive values of time delay $\tau$. As it can be appreciated from the figure, a marked cross-correlation peak, centered in $\tau=0$, is present. The characteristic width of this peak varies between  $5$ and $20$ minutes, depending on the cell line and on the age.
The fairly symmetric shape of the cross-correlation peak with respect to the vertical axis indicates that there is no systematic delay between divergence and nuclear deformation, at least within the experimental temporal resolution ($2$ minutes).

The evidence of an instantaneous correlation between $\dot{a}_j\left(t\right)$ and $\nabla\cdot\Vec{v}_j\left(t\right)$ prompts us to further investigate it to understand its character and origin.
For each one of the $8$ hours time intervals, we therefore generate a scatter plot of $\dot{a}_j\left(t\right)$ versus  $\nabla\cdot\Vec{v}_j\left(t\right)$. To make the plot more readable, we introduced a uniform binning of the horizontal axis. For each bin, containing on average hundreds or thousands of points of the scatter plot, we report in figure \ref{Jam_DCIS_HC}.c the average value $\dot{a}_n$ of the strain rate as a function of the average value $\nabla\cdot\Vec{v}$ of the divergence. Remarkably, at least in the vicinity of the origin, the obtained curves display a fairly linear behavior with zero intercept, thus denoting a direct proportionality between the nucleus strain rate and the corresponding divergence of the velocity field the nucleus is subjected to. In order to rationalize the physical meaning of the proportionality constant $m$ between $\nabla\cdot\Vec{v}$ and $\dot{a}_n$, we note that, under the hypothesis of weak spatial dependence of the monolayer cell number density $\rho_{c}$, the continuity equation for the monolayer density can be written as

\begin{equation}
    \nabla\cdot\Vec{v} = -\dot{\rho_{c}} = \dot{a}_{c},
\end{equation}

where $a_c$ is instantaneous cell area divided by its mean value and $\dot{a}_c$ is the cell strain rate.
The coefficient $m$ can be therefore interpreted as the ratio between the nuclear area strain rate and the cell area strain rate. The reciprocal value $e_n = m^{-1}$ is related to the relative stiffness of the nucleus compared to the one of the entire cell: the larger is $e_n$, the lower is the deformation per unit time of the nucleus compared to the one of the cell. Estimates of $m$ for the different cell lines and ages are obtained \textit{via} a linear fit of the corresponding $\dot{a}_n$ versus $\nabla\cdot\Vec{v}$  in the neighborhood of the origin.
The corresponding values of $e_n$ are reported in figure \ref{Jam_DCIS_HC}.f.
Interestingly enough, the relative stiffness has a  time dependence which is rather different from the one displayed by both the divergence and the strain rate. While in all jamming monolayers both $\sigma_\nabla$ and $\dot{a}_0$ reach a stable plateau value after about $12$ hours, the relative stiffness typically keep on decreasing over time, with a roughly constant rate.
As it can be appreciated from figure \ref{Jam_DCIS_HC}.c, for large values of $\nabla\cdot\Vec{v}$ a significant deviation from linearity is observed, with $\dot{a}_n$ seemingly approaching a saturation value for both negative and positive strain rates. The ratio between nuclear strain rate and cell strain rate becomes thus lower and lower as the imposed  strain rates increase. 

\subsection{Flocking monolayers}
\subsubsection{In flocking monolayers, nuclei experience stronger deformations and are stiffer}
\begin{figure*}
  \includegraphics[width=\textwidth]{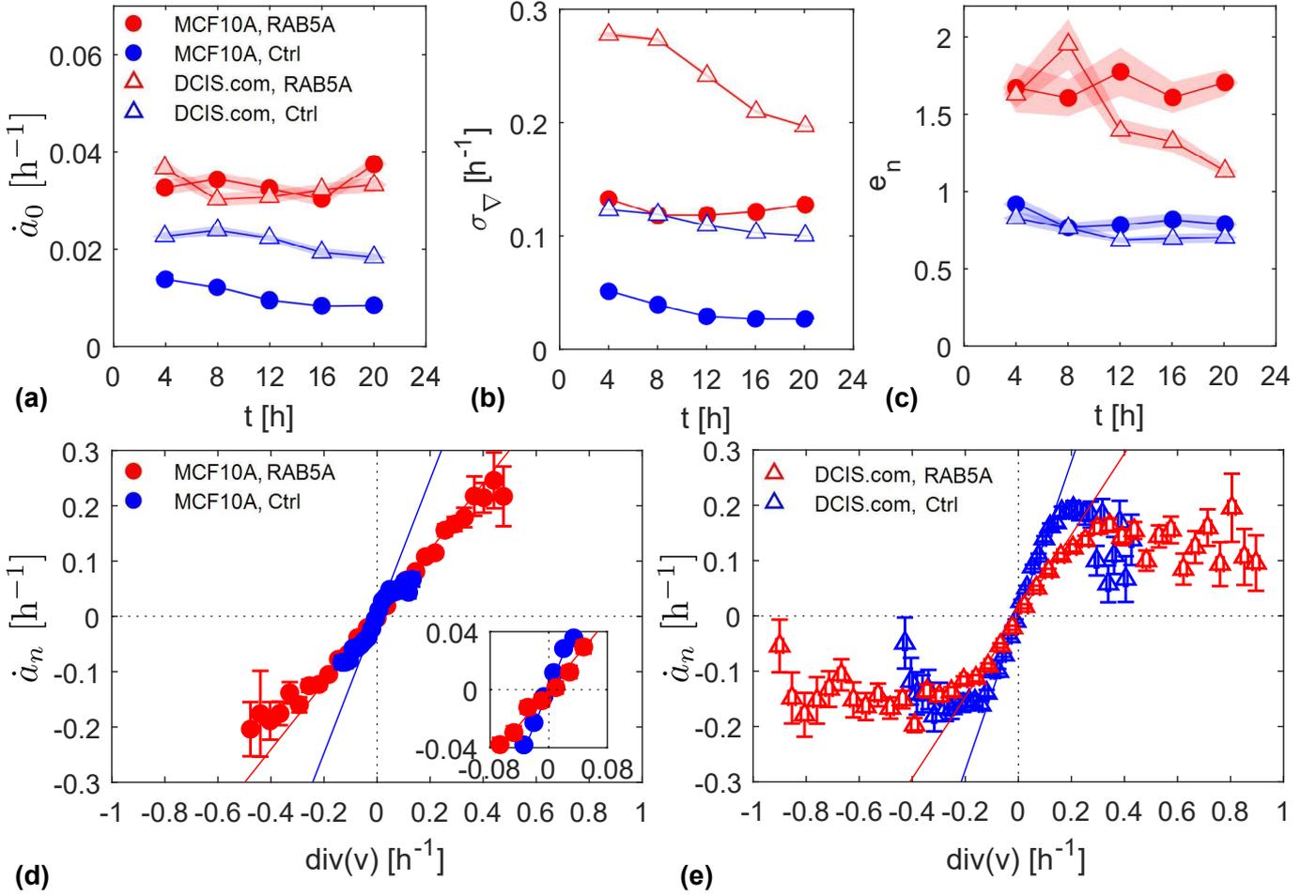}
\caption{\label{fig_stiff_flocking}
Data relative to control (blue) and RAB5A overexpressing (red) cells in experiments on flocking monolayers. Results relative to MCF10A (plain circles) and MCF10A.DCIS.com (void triangles) are presented.
\textbf{a.} Nuclear characteristic strain rates $\dot{a}_0$ as a function of time, obtained from the fit of the corresponding MSS relative to the jamming experiments.
\textbf{b.} Divergence standard deviation as a function of time. 
\textbf{c.} Reciprocal $e_n$ of the slope of $\dot{a}_n$ versus $\nabla\cdot\Vec{v}$ scatter plot, obtained from linear fit of the linear part of the scatter plots.
\textbf{d,e.}  Scatter plot of the instantaneous single nucleus strain rate $\dot{a}_n$ as a function of the corresponding divergence. Data relative to MCF10A (d) and MCF10A.DCIS.com (e) monolayers. Reported points are average over hundreds/thousands of strain rate-divergence couples made over bins equally spaced along the abscissae axis. Error bars are evaluated as standard deviations of the mean of the data contained in the bins. Lines are linear fits with zero intercept made over the linear part of the scatter plots. In the inset of (d), it is reported a zoom of the scatter plot over the area where both samples exhibit a linear behaviour.
}
\end{figure*}

To further investigate the impact of cell motility on nuclear deformation, we repeated the  experiments using two different cell lines (MCF10A and MCDF10A.DCIS.com) whose dynamical state is perturbed by inducing the overexpression of the RAB5A protein, a master regulator of endocythosis \cite{zerial2001rab}. 
Upon RAB5A overexpression, mature, almost completely kinetically arrested monolayers experience a dramatic reawakening of motility, characterized by highly coordinated directed migration (flocking) and by the presence of local cell rearrangments (fluidization). \cite{malinverno2017endocytic,giavazzi2018flocking,giavazzi2017giant}.
For both considered cell lines, we compare RAB5A overexpressing monolayers with the corresponding controls in order to investigate the impact of the flocking transition on nuclear deformation. As detailed in appendix \ref{appe_flocking}, PIV analysis confirms the striking motility phenotype of RAB5A overepressing monolayer previously reported \cite{malinverno2017endocytic,palamidessi2019unjamming}. In figure \ref{fig_stiff_flocking} we show  the average nuclear strain rate (fig. \ref{fig_stiff_flocking}.a) and the standard deviation of the divergence of the velocity field (\ref{fig_stiff_flocking}.b) for the two cell lines. As it can be seen, both $\dot{a}_0$ and $\sigma_{\nabla}$ are systematically larger in flocking monolayers, where the relative motion of cells is enhanced (leading to larger values of divergence) and, as a consequence, nuclei are subjected to larger stresses and deform more. Despite the significant difference in $\dot{a}_0$ and $\sigma_{\nabla}$ between the two cell lines, a striking correspondence between the two series of experiments can be observed in the slopes of the linear regime of the scatter plots reported in figure \ref{fig_stiff_flocking}.d and \ref{fig_stiff_flocking}.e: as it can be clearly seen in figure \ref{fig_stiff_flocking}.c, values of $e_n$ are very similar for the corresponding  conditions, with a systematic difference by a factor $\sim2$ between the RAB5A overexpressing monolayers and the control ones. This suggests that the nuclei of RAB5A overexpressing cells deform less, compared to the control ones, when subjected to the same local density variation. 

The $\dot{a}_n$ versus $\nabla\cdot\Vec{v}$ curves reported in panels \ref{fig_stiff_flocking}.d,e display some interesting features even beyond the linear regime. As far as MCF10A cells are concerned, control monolayers display a clear deviation from the linear regime when $|\dot{a}_n|$ is above $\sim 0.05$ h\textsuperscript{-1}, while the corresponding curve for RAB5A overexpressing cells remains linear for $|\dot{a}_n|<0.1$ h\textsuperscript{-1}. On the other hand, for MCF10A.DCIS.com cells, both control and RAB5A overexpressing monolayers display a non-linear behaviour at nuclear strain rates larger than $\sim 0.1$ h\textsuperscript{-1} and, apparently, the nuclear strain rate seems to attain a plateau at $|\dot{a}_n|\sim0.2$ h\textsuperscript{-1}. The occurrence of such a plateau suggests the presence of mechanisms preventing the nuclei to be deformed above a certain threshold. Moreover, the marked difference between the two cell lines is even more interesting considering that MCF10A.DCIS.com is a tumoral cell line deriving from MCF10A \cite{millerMCF10DCIS}. 
Further investigation is required to clarify whether the expanded  linear regime (and the wider dynamic range of $\dot{a}_n$) displayed by  MCF10A.DCIS.com control monolayers compared to the corresponding MCF10A ones could be related to the tumorogenic potential of the cells.

\subsection{Model}

\begin{figure}
\resizebox{1\columnwidth}{!}{%
  \includegraphics{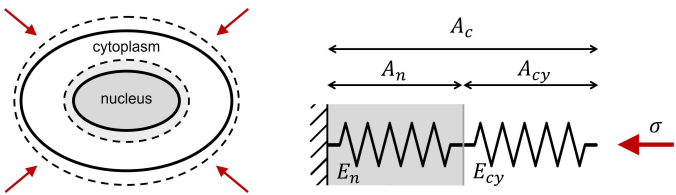}}
\caption{\label{fig_sketch_spring}
Schematic representation of a cell deforming under the action of an in-plane compressive stress. On the right, 1D model of the cell as a series of two elastic elements with different \textcolor{black}{elastic} moduli $E_n$ and $E_{cy}$, corresponding to the nucleus and the cytoplasm, respectively. Corresponding areas $A_n$ and $A_{cy}$ are represented as lengths of the springs in the 1D model. The total area $A_c$ given by the sum of $A_n$ and $A_{cy}$.}
\end{figure}


\textcolor{black}{Our in-plane observations do not allow discriminating whether the observed fluctuations in nuclear and cellular projected area can be assumed to occur at constant volume, with each change in the projected area corresponding to a height variation of opposite sign. In principle, our observations are compatible with an actual variation of cell and nuclear volume, mediated by fluid exchange between nucleus and cytoplasm and extracellular medium. Large volume fluctuations
in multicellular aggregates, due to active water transport in and
out of the cell have been indeed reported.  The complex interplay between single-cell volume fluctuations and collective monolayer dynamics has been investigated \textit{via} a combination of velocimetry and single-cell segmentation, similar to one exploited also in the present work \cite{zehnder2015multicellular,zehnder2015cell}.
Although the reported timescale of these volume fluctuations (a few hours) \cite{zehnder2016langevin} is typically much larger than the correlation time of the projected area fluctuations observed in our experiments (a few minutes), we cannot rule out the possibility that the observed behavior is actually due to a poroelastic-like response of the cell \cite{darcy1856fontaines}. 
Even without making assumptions on the details of the underlying process, we can provide a description of the mechanical properties of cells and nuclei in terms of effective quantities, building on the experimentally accessible observables described in the previous sections.}
Given the limited temporal resolution of our experiments (2-10 minutes) we do not expect to be able to capture the full rheological response of the cells. In particular, we can hardly probe the process of viscoelastic relaxation upon stress application, which typically occurs on the scale of few tens of seconds \cite{khalilgharibi2019stress}. 
\textcolor{black}{We are then most likely sensitive to the elastic response of cell components.}
Indeed, the lack of delay between $\dot{a}_n$ and $\nabla\cdot\Vec{v}$ observed from their cross-correlation (see figure \ref{fig_divdacorr}) \textcolor{black}{is consistent with this hypothesis.} 

We introduce a simple phenomenological model (sketched in figure \ref{fig_sketch_spring}) \textcolor{black}{for the mechanical response of the cell to in-plane homogeneous compressive or tensile stresses}, where an elastic nucleus and an elastic cytoplasm, with constant \textcolor{black}{elastic} moduli $E_n$ and $E_{cy}$, respectively, are connected in series 
Within this simple model, we can derive an explicit relationship between the ratio $E_n/E_{cy}$ of the elastic moduli and relative stiffness $e_n$ introduced in section \ref{section_corrVelNuc}. 
For two elastic components in series, the total area deformation $\Delta A_{c}$ is the sum of the deformations $\Delta A_{n}$ and $\Delta A_{cy}$ of the components, while the external \textcolor{black}{} stress $\sigma$ applied on the system coincides with the one applied on each element

\begin{equation}
\sigma\propto\left(E_n a_n\right) = \left(E_{cy} a_{cy}\right).   
\label{stressstrain}
\end{equation}

By introducing the ratio between the average projected nuclear area and the average projected cell area $\beta\doteq\left<A_n\right>/\left<A_{c}\right>$, it is possible to write the total strain associated to the cell projected area $a_{c}=\Delta A_{c}/A_{c}$ in terms of the nuclear $a_n$ and the cytoplasmic strain $a_{cy}$, as

\begin{equation}
    a_{c} = \beta a_n + \left( 1 - \beta \right) a_{cy}.
\label{eq_strainbeta}
\end{equation}

Combining this equality with the time derivative of equation \ref{stressstrain}, we get

\begin{equation}
    \frac{E_n}{E_{cy}}= \frac{e_n-\beta}{1-\beta}
    \label{eq_modratio}
\end{equation}

\begin{figure}
\resizebox{1\columnwidth}{!}{%
  \includegraphics{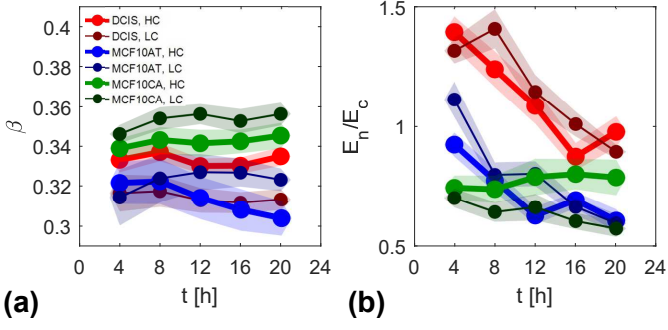}}
\caption{\label{fig_beta_e_mod}
Time evolution of the surface fraction $\beta$ covered by the nuclei (\textbf{a.}) and of the \textcolor{black}{elastic} moduli ratio $E_n/E_{cy}$ (\textbf{b.}) for the set of experiments on jamming monolayers. Data reported as a function of time in the same time intervals of figures \ref{Jam_DCIS_HC}.d-f. Error bars evaluated by propagating the corresponding errors.
}
\end{figure}

A time-resolved estimate of 
$\beta$ can be obtained directly from the images, by combining the average nuclear area (Fig. \ref{fig_A_si}) with the average cell area, which can be simply estimated as the total area of the FOV divided by the number of identified cells. In figure \ref{fig_beta_e_mod}.a we report the resulting time evolution of $\beta$ for the jamming monolayers, evaluated over the same time intervals considered in panels \ref{Jam_DCIS_HC}.d-e-f. We note that $\beta$ takes slightly  different values for different samples, and it is approximately time independent.

Once $\beta$ is known, we can use equation \ref{eq_modratio} to obtain an estimate of the  \textcolor{black}{ratio between the elastic moduli} 
As it can be seen from figure \ref{fig_beta_e_mod}.b, DCIS moduli ratio uniformly decreases from values above unity to values less than $1$. Their nuclear modulus is therefore initially larger than the one of chytoplasm, but the relation reverses as far as cell packing increases. Similarly, in MCF10AT initially both nucleus and cytoplasm have similar moduli but with time cytoplasm becomes stiffer than nucleus. Finally, in MCF10CA $E_{cy}$ is systematically larger than $E_n$. Possibility to observe nuclei either with lower or higher \textcolor{black}{elastic} modulus is not surprising as already observed in literature \cite{fischer2020effect} 
Robustness of the presented results is granted by the small relative errors, obtained from the standard deviation of the mean evaluated over $5$ different FOVs for each sample, and by the similarity of values and trends among difference seeding concentrations of same samples.

\begin{figure}
\resizebox{1\columnwidth}{!}{%
  \includegraphics{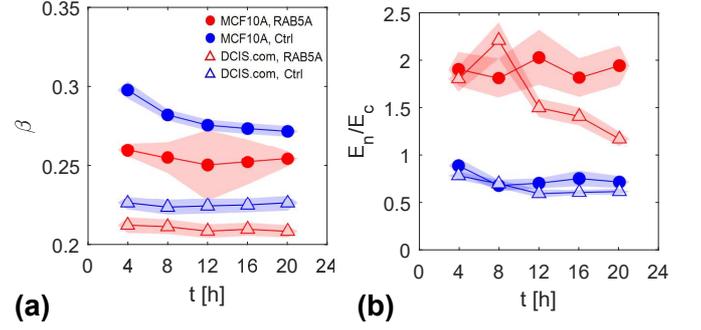}}
\caption{\label{fig_beta_e_mod_RAB5}
Time evolution of the surface fraction $\beta$ covered by the nuclei (\textbf{a.}) and of the \textcolor{black}{elastic} moduli ratio $E_n/E_{cy}$ (\textbf{b.}) for the set of experiments on flocking monolayers. Data reported as a function of time in the same time intervals of figures \ref{fig_stiff_flocking}.a-c. Error bars are evaluated by propagating the corresponding errors.
}
\end{figure}

Through the same procedure, we estimate $\beta$ (fig. \ref{fig_beta_e_mod_RAB5}.a) and $E_n/E_{cy}$ (fig. \ref{fig_beta_e_mod_RAB5}.b) for the flocking monolayers. While in control samples the \textcolor{black}{elastic} modulus of the nuclei is always close to the one of the cytoplasm, in RAB5A overexpressing cells it is almost twice as large as the cytoplasmic one, for both MCF10A and MCF10A.DCIS.com. In principle, these results would be compatible with both nuclear stiffening and cytoplasmic softening, as a consequence of RAB5A overexpression. However, AFM measurements of monolayer rigidity (which are known to probe mainly nuclear stiffening) have already documented in MCF10A cell line an increase by a factor two in the \textcolor{black}{Young's} modulus in RAB5A overexpressing confluent cells \cite{malinverno2017endocytic}.   


\section{Conclusions}

In this work, we have introduced and demonstrated an experimental non-invasive procedure aimed at extracting information on the relative stiffness of nuclei compared to the cytoplasm, for epithelial cell monolayers whose state is monitored during time-lapse microscopy experiments. The procedure involves segmentation and tracking of the fluorescently-tagged nuclei, combined with PIV and PT analysis of the nuclear motion.

More specifically, space-resolved and time-resolved studies of the monolayer dynamics and the nuclear deformation allows the study of their interplay, from which we find that for small strain rates, a linear correlation holds between the local density changes, estimated via the divergence of the velocity field, and the nuclear strain rate.
\textcolor{black}{In this linear regime, consistent with an elastic response,}\textcolor{blue}{[In this linear elastic regime,]} we extract information on the \textcolor{black}{effective} relative stiffness of the nuclei, as we show with experiments performed with both jamming and flocking monolayers. For jamming monolayers of three different cell lines, we find nuclear moduli close to the ones of cytoplasm; we also find that, as the monolayers jam over time and the dynamics lowers and internal agitation within the monolayer decreases, the nucleus softens compared to the cytoplasm. In agreement with this observation, a set of experiments on flocking monolayers obtained by overexpression of the RAB5A protein in two cell lines shows that the nuclear relative stiffness is higher than in control samples, despite the fact that nuclei deform more. The difference in the relative nuclear moduli is significant (about a factor of two) and is surprisingly similar for the two cell lines.

For large strain rates, we observe nuclear stiffening, with strain rates eventually approaching a plateau as a function of the divergence of the velocity field. We interpret this result as a mechanoprotective response of the nuclei to oppose large stresses that arise from the increased internal agitation of the nuclei in a highly dynamic monolayer. Interestingly, our results are compatible with the view that the value of the nuclear \textcolor{black}{elastic} modulus for small strains is mainly determined by chromatin, whereas for larger strains the strain-stiffening contribution of lamins dominates \cite{stephens2017chromatin,stephens2018separate}.  

The possibility of non-invasive monitoring of the relationship between mechanical stimuli, nuclear deformations and nuclear relative stiffness in cell monolayers may have immediate biological applications. An early version of the methodology that we describe here was recently used in Ref. \cite{frittoli2021tissue} to assess in real time the nuclear mechanical stress and response and the associated role in causing a long-term transcriptional-dependent phenotype as a consequence of the short-term adaptive response to stress. We thus believe that our methodology represents a useful addition to the portfolio of non-invasive tools to characterize the mechanobiological response of tissues.

\section*{Acknowledgements}
This work has been supported by: the Associazione Italiana per la Ricerca sul Cancro (AIRC) to F.G. and S.V. (MFAG\#22083), and to G.S. (IG\#18621 and 5XMille \#22759); the Italian Ministry of University and Scientific Research (MIUR) to G.S. (PRIN: Progetti di Ricerca di Rilevante Interesse Nazionale – Bando 2017\#2017HWTP2K).

\section*{Author contribution statement}
F.G. conceived research. R.C., G.S. and F.G. designed research. A.P. and E.F. prepared the samples and performed the experiments. S.V. developed the image processing pipeline, wrote the codes and analyzed all the data. S.V., R.C. and F.G. wrote the manuscript. All Authors contributed discussing the results and the manuscript.
\appendix

\section{. Mean square displacement analysis}
\label{appe_MSD}
\begin{figure*}
  \includegraphics[width=\textwidth]{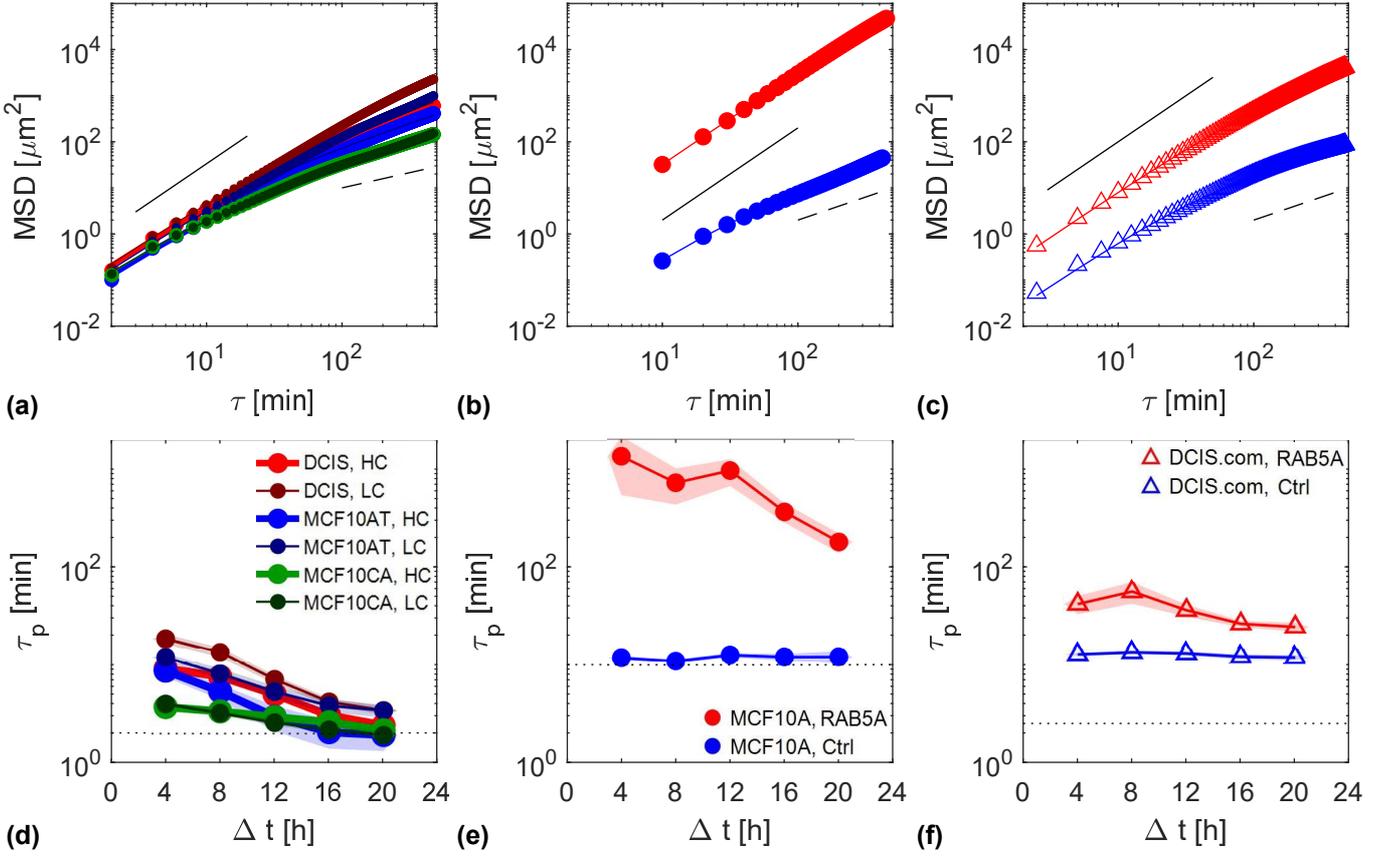}
\caption{\label{fig_MSD}
\textcolor{black}{First row: nuclear mean square displacements (MSD) obtained from experiments on jamming monolayers (\textbf{a.}), on MCF10A CTRL and RAB5A overexpressing flocking monolayers (\textbf{b.}), and on DCIS.com CTRL and RAB5A overexpressing flocking monolayers (\textbf{c.}). In the legend of panel (\textbf{a.}) HC and LC correspond to high and low seeding density, respectively. The reported MSD are averaged over the entire duration of the experiment. Corresponding best fitting curves with a ballistic-to-diffusive transition model (see Eq. \ref{transition}) are reported as lines of the same colors. Black continuous and dashed lines represent guidelines for ballistic and diffusive dynamics, respectively. Each curve is plotted after subtraction of a constant offset  $\sigma_{MSD}$, obtained from the fitting procedure (see Eq. \ref{transition}). In the second row, we display the persistence times obtained for the jamming experiments (\textbf{d.}), the MCF10A flocking experiment (\textbf{e.}), and the DCIS.com flocking experiment (\textbf{f.}), within each one of the five considered time windows. In each panel, the dotted line correspond to the value $\delta t$ of the delay time between two consecutive acquired frames.
}
}
\end{figure*}

\textcolor{black}{In this Appendix, we discuss the mean square displacements (MSDs) associated to the trajectories obtained from nuclear tracking, as described in section \ref{tracking}. For each set of experiments, the MSD is calculated within each of the five time windows considered in the main text. In figure \ref{fig_MSD}.a-b-c, we display averages over all five time windows as a function of the lag time $\tau$. There, continuous and dashed black lines indicates the slope corresponding to a ballistic scaling (quadratic in $\tau$), and a diffusive one (linear in $\tau$), respectively. As it can be appreciated also visually, in all experiments a crossover from a ballistic-like scaling at short time delays to a diffusive-like scaling  at larger time scales is observed.
The characteristic time $\tau_p$ marking the transition between the two regimes, which corresponds to the persistence time during which the velocity of a given cell is constant, can be estimated by fitting to the data the function}

\begin{equation}
\label{transition}
    MSD\left(\tau\right) = \sigma_{MSD} + 2\tau_p v_b^2 \left[\tau + \tau_p\left(e^{-\tau/\tau_p}-1\right)\right]
\end{equation}

\textcolor{black}{where $v_b$ is the ballistic velocity and $\sigma_{MSD}$ is an offset to account for experimental noise \cite{uhlenbeck1930theory}. The persistence times recovered from the fit are reported in figure \ref{fig_MSD}.d-e-f. As it can be seen, they are systematically larger than the time step between two acquired frames in the corresponding experiments, represented by horizontal dotted lines. This observation implies that, on the timescale $\delta t$ corresponding to the delay between two consecutive images, cells are moving with good approximation with constant velocity. This confirms the validity of the assumption made in section \ref{sec_dynpar} that the instantaneous nuclear velocity can be reliably estimated as the average velocity between two consecutive frames.}

\section{. Comparison of PIV results from fluorescent and phase contrast time lapses}
\label{appe_PIV}

In this Appendix, we compare the dynamics parameters obtained from PT and from PIV made over fluorescent and phase contrast time lapses. We then deepen the discrepancies between the velocities and velocity divergences recovered from PT and from PIV analysis of fluorescent and phase contrast time lapses, in order to motivate the choice of the fluorescent PIV analysis to evaluate the divergence of the velocity field as an indicator of the local density fluctuations within the monolayer.

\begin{figure}
\resizebox{1\columnwidth}{!}{%
  \includegraphics{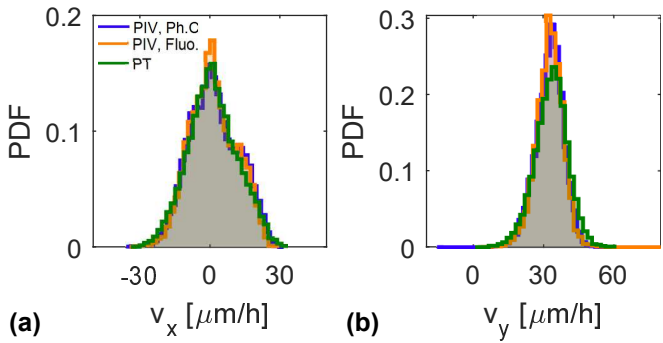}}
\caption{\label{fig_PIV-PT_histo}
Velocity components $v_x$ (\textbf{a.}) and $v_y$ (\textbf{b.}) distribution within an MCF10A RAB5A overexpressing FOVs. Distributions are obtained from PIV computed on phase contrast images (blue), from PIV computed over fluorescent images (orange) and from PT (green). Distribution are recovered in the time interval $8$-$9$ h.
}
\end{figure}

In figure \ref{fig_PIV-PT_histo} are reported the velocity distributions along the two main axis for a flocking MCF10A RAB5A overexpressing monolayer. The reported dynamics refers to the $8$-$9$ hours time interval, corresponding to the cell motility peak \cite{malinverno2017endocytic}. Blue, orange and green lines refer to PIV computed over phase contrast and fluorescent time lapses and to PT, respectively. The recovered distributions are very similar to each other, thus validating the obtained dynamics.


\begin{figure}
\resizebox{1\columnwidth}{!}{%
  \includegraphics{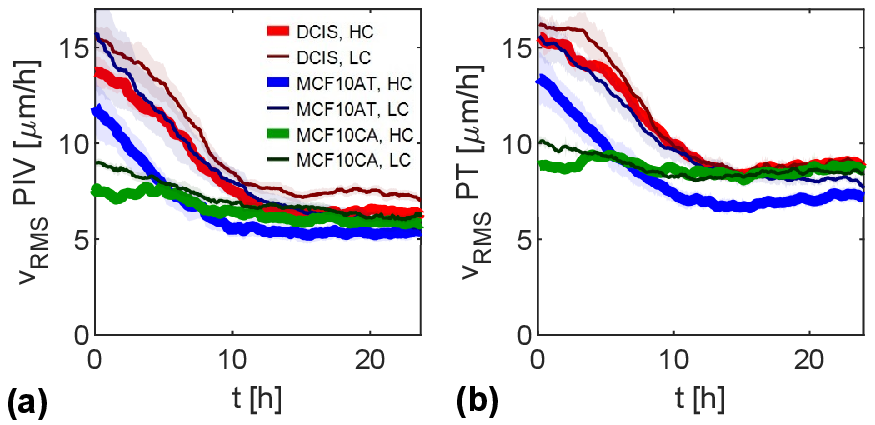}}
\caption{\label{fig_vrmsPIV}
Time evolution of RMS velocity ${v}_{RMS}$ obtained from PIV on fluorescent time lapses (\textbf{a.}) and PT (\textbf{b.}). Shadowed error bars are avaluated as the standard deviation of the mean evaluated for each sample over the 5 different FOVs. Reported data refer to jamming monolayers: DCIS HC (red thick line) and LC (dark red thin line), for MCF10AT HC (blue thick line) and LC (dark blue thin line) and for MCF10CA HC (green thick line) and LC (dark green thin line).}
\end{figure}

Mean values obtained from PT and PIV are also very close, as shown in figure \ref{fig_vrmsPIV} where are reported for the six jamming experiments the time evolution of $v_{RMS}$ as recovered from PIV made over fluorescent time lapses (fig. \ref{fig_vrmsPIV}.a) and from PT (fig. \ref{fig_vrmsPIV}.b).

Given the correspondence between the results obtained from PT and from PIV, it is possible to choose which of the two methods to use to compare velocity divergence with the nuclear deformations shown in section \ref{section_corrVelNuc}. 
A first possibility would be to start from the single nuclei velocities $\Vec{v}_j\left(t_i\right)$ to build up frame by frame a velocity field through a proper interpolation of the known velocities in the plane. This choice has been discarded in order to avoid spurious correlations deriving from the common origin from the localisation and segmentation of nuclei of the correlating data (divergence of the velocity field and time derivative of the area). We therefore chose to use a divergence obtained from PIV to ensure the authenticity of any correlation. Additionally, even a 10\% missed nuclei in PT while allowing a fair enough agreement with PIV results in the velocity distribution (Fig. \ref{fig_PIV-PT_histo}), at the same time prevents a proper interpolation of the velocity field. Test made revealed a discrepancy even by a factor 2 in the width of the $\nabla\cdot\Vec{v}$ distributions between PT and PIV, either from fluorescent and phase contrast analysis.

\begin{figure}
\resizebox{1\columnwidth}{!}{%
  \includegraphics{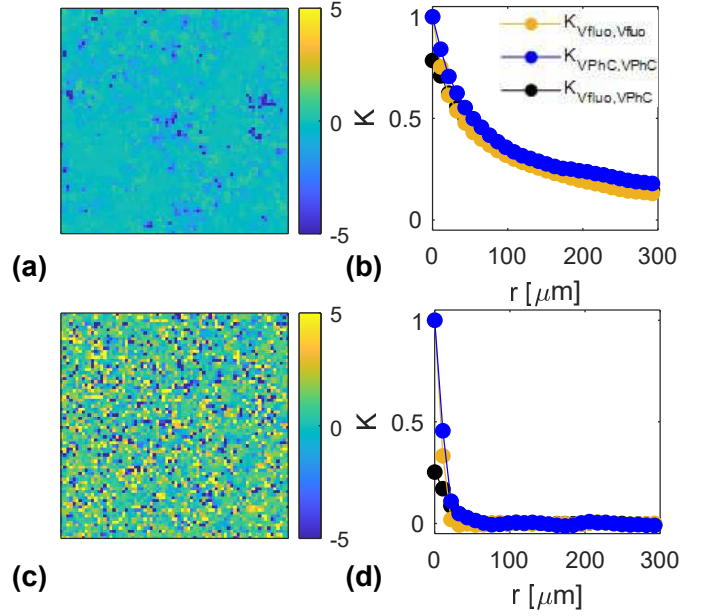}}
\caption{\label{fig_appe_PIV_corr}
\textbf{a.} Relative difference between the modulus of the velocity field obtained from PIV from fluorescent and phase contrast images.
\textbf{b.} Azimuthal average of the space correlations $K_{v_{Fluo},v_{Fluo}}$ (orange), $K_{v_{PhC},v_{PhC}}$ (blue) and $K_{v_{Fluo},v_{PhC}}$ (black).
\textbf{c.} Relative difference between the divergence of the velocity field obtained from fluorescent and phase contrast images.
\textbf{d.} Azimuthal average of the space correlations  $K_{\nabla_{Fluo},\nabla_{Fluo}}$ (orange), $K_{\nabla_{PhC},\nabla_{PhC}}$ (blue) and $K_{\nabla_{Fluo},\nabla_{PhC}}$ (black).}
\end{figure}

A deeper analysis revealed that even between the velocity fields recovered applying PIV analysis on fluorescent and phase contrast images have some discrepancies. We evaluate the normalized space cross-correlation between the velocity fields moduli $v_{Fluo}\left(\Vec{x}\right)$ and $v_{PhC}\left(\Vec{x}\right)$ obtained from the PIV analysis of the two data images respectively:

\begin{equation}
   K_{v_{Fluo},v_{PhC}}\left(\Vec{r}\right) =  \frac{\left< v_{Fluo}\left(\Vec{x}\right) \cdot v_{PhC}\left(\Vec{x}+\Vec{r}\right) \right>_{\Vec{x}} } {\sqrt{\left<v_{Fluo}\left(\Vec{x}\right)\right>_{\Vec{x}}^2 \left<v_{PhC}\left(\Vec{x}\right)\right>_{\Vec{x}}^2}}
\end{equation}

where $\left<...\right>_{\Vec{x}}$ stands for the spatial averages over all the positions where the velocity fields are evaluated. In figure \ref{fig_appe_PIV_corr}.b it is reported the azimuthal average of $K_{v_{Fluo},v_{PhC}}$ together with the space self-correlations of $v_{Fluo}\left(\Vec{x}\right)$ and $v_{PhC}\left(\Vec{x}\right)$. The spatial scale of decorrelation is similar in the three cases, but the cross-correlation value in $r=0$ is lower by $20$\% than the self-correlations, thus testifying a weak but existent difference in the velocity fields.

Similarly, in figure \ref{fig_appe_PIV_corr}.d we report the azimuthal average of the cross-correlation relative to the divergences of the velocity fields defined as: 

\begin{equation}
   K_{\nabla_{Fluo},\nabla_{PhC}}\left(\Vec{r}\right)] =  \frac{\left< \left[\nabla\cdot\Vec{v}_{Fluo}\left(\Vec{x}\right)\right] \left[\nabla\cdot\Vec{v}_{PhC}\left(\Vec{x}+\Vec{r}\right) \right>_{\Vec{x}}\right] } {\sqrt{\left<\nabla\cdot\Vec{v}_{Fluo}\left(\Vec{x}\right)\right>_{\Vec{x}}^2 \left<\nabla\cdot\Vec{v}_{PhC}\left(\Vec{x}\right)\right>_{\Vec{x}}^2}}
\end{equation}

As in the previous case, self-correlations of $\nabla\cdot\Vec{v}_{Fluo}\left(\Vec{x}\right)$ and $\nabla\cdot\Vec{v}_{PhC}\left(\Vec{x}\right)$ are reported in the same plot.
For the divergences, the cross-correlation in $r=0$ is much lower than the correspondent self-correlations, thus pointing out much larger differences between the divergence fields evaluated with the two methods.
Discrepancy between the fields of $\nabla\cdot\Vec{v}$ much more  pronounced than the ones in the corresponding $v$ can also be appreciated comparing figures \ref{fig_appe_PIV_corr}.a and \ref{fig_appe_PIV_corr}.c, where the relative difference between the fluorescent and the phase contrast analysis are represented for the modulus of the velocity field and the divergence respectively. While for $v$ the discrepancies only emerges in the regions where there is a low density of nuclei (dark blue regions in fig. \ref{fig_appe_PIV_corr}.a), very different appear to be $\nabla\cdot\Vec{v}_{fluo}$ from $\nabla\cdot\Vec{v}_{phc}$. We note that such differences are only present when spatial resolution is considered. Distributions of the divergence, frame by frame, are very close in the two cases.

\begin{figure}
\resizebox{1\columnwidth}{!}{%
  \includegraphics{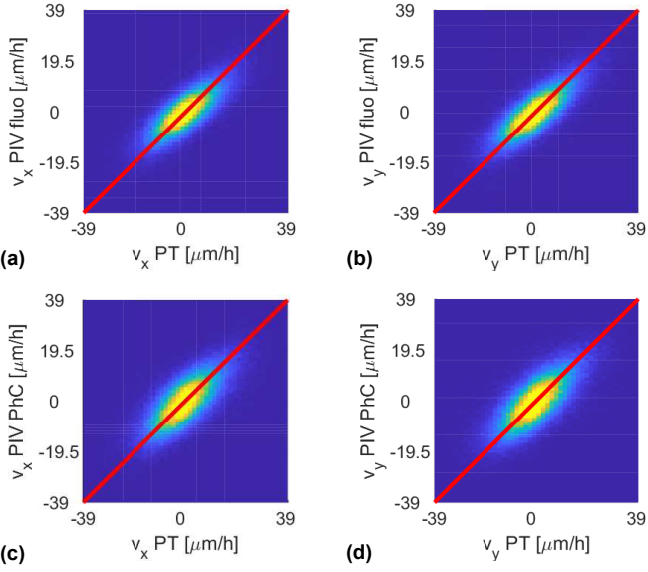}}
\caption{\label{fig_appe_PIV_v}
\textbf{a-b.} Correlation between the velocities components recovered from PT and the ones obtained interpolating at the nuclei center coordinates the velocity field obtained from PIV on fluorescent time lapses. The minor widths of the distributions, evaluated as the lower eigenvalue of the covariance matrix, are $\sigma_x=0.012$ and $\sigma_y=0.012$ for $v_x$ and $v_y$ respectively.
\textbf{c-d.} Correlation between the velocities components recovered from PT and the ones obtained interpolating at the nuclei center coordinates the velocity field obtained from PIV on phase contrast time lapses. The minor widths of the distributions are now $\sigma_x=0.016$ and $\sigma_y=0.017$ for $v_x$ and $v_y$ respectively.
Scatter plots are made considering the frames in the time interval $1$-$8$ hours relative to one of the field of view of DCIS HC jamming experiments.}
\end{figure}

Given the differences, a choice has to be made between the two methods for the comparison with the nuclear strain rates. To this aim, we report in figure \ref{fig_appe_PIV_v} scatter plots of the single nuclei velocity components $v_{x,j}$ (fig.  \ref{fig_appe_PIV_v}.a,c) and $v_{y,j}$ (fig.  \ref{fig_appe_PIV_v}.b,d) obtained from PT as a function of the corresponding components of the cubic interpolation of the velocity field $\Vec{v}_{fluo}$ (fig.  \ref{fig_appe_PIV_v}.a,b) and $\Vec{v}_{phc}$ (fig.  \ref{fig_appe_PIV_v}.c,d) in the center of mass of the same nuclei. Looking at dispersions, one can find out that correspondence is better between PT and PIV analysis of fluorescent time lapses. This can be more easily sensed by looking at the width of the distributions along their minor dimension. This can be made by compuing the minor eigenvalue of the covariance matrix between the velocity component obtained from PT and from PIV. The resulting widths are $\sigma_x = 0.012$ and $\sigma_y=0.012$ for the fluorescent time lapses case and $\sigma_x = 0.016$ and $\sigma_y=0.017$ for the phase contrast one.

Larger discrepancies from PT in the latter case can be partially rationalized considering that in phase contrast inner dynamics of the cells is also visible. Moreover, visual inspection of the videos revealed that several bright spots appear in phase contrast localized at the edges between adjacent cells. They move along cell edges faster than the cells themselves, thus affecting the velocity obtained from PIV. 

In conclusion, because of this larger discrepancy between PT and phase contrast, we choose to compare area strain rate with the divergence obtained from PIV analysis of fluorescent time lapses.

\section{. Self and Cross correlations between local divergence and nuclear strain rate}
\label{appe_corr}

In the main text, we report the normalized cross-correlation between nuclear strain rate $\dot{a}_j\left(t\right)$ and the corresponding divergence of the velocity field $\nabla\cdot\Vec{v}_j$. Normalized cross-correlation for generic variables $q_j\left(t\right)$ and $p_j\left(t\right)$ is defined as

\begin{equation}
    K_{p,q}\left(\tau\right) = \frac{\left<\left<p_j\left(t+\tau\right) \cdot q_j\left(t\right)\right>_t\right>_j}{\sqrt{\left<\left<p_j^2\left(t\right)\right>_t\right>_j^2 \cdot \left<\left<q_j^2\left(t\right)\right>_t\right>_j^2}}
\end{equation}

\begin{figure}
\resizebox{1\columnwidth}{!}{%
  \includegraphics{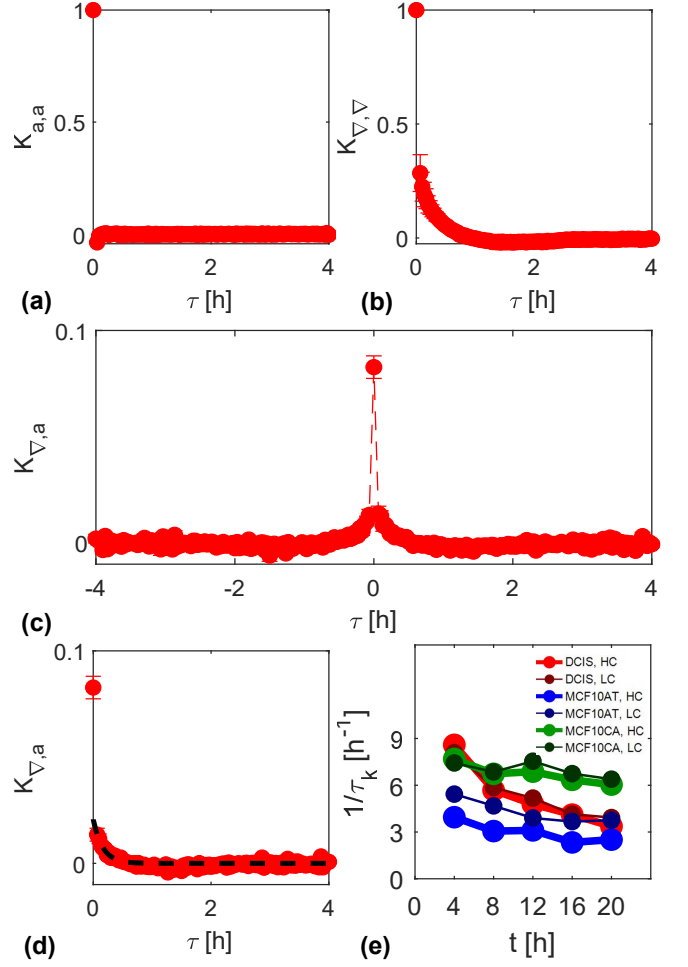}}
\caption{\label{fig_divdacorr}
\textbf{a.} Normalized autocorrelation of the instantaneous nuclear strain rate $\dot{a}_n$. Data relative to jamming experiment DCIS HC in the time interval $0$-$24$h.
\textbf{b.} Normalized autocorrelation of the divergence of the velocity field. Data relative to jamming experiment DCIS HC in the time interval $0$-$24$h.
\textbf{c.} Normalized cross-correlation between the instantaneous nuclear strain rate $\dot{a}_n$ and the divergence of the velocity field evaluated in the center of mass of the same nucleus. Data relative to jamming experiment DCIS HC in the time interval $0$-$24$h. 
\textbf{d.} Same data of c. but averaging over positive and negative $\tau$. Black dashed line represent the best exponential fit detailed in the text.
\textbf{e.} Decay rate of $K_{\dot{a},\nabla}$ as a function of the considered time interval for the different experiments on jamming monolayers. Red, blue and green refer to DCIS, MCF10AT and MCF10CA monolayers, respectively. Thin and thick lines correspond to monolayers seeded at low and high concentration, respectively.
}
\end{figure}

In figure \ref{fig_divdacorr} are reported the nuclear strain rate (fig. \ref{fig_divdacorr}.a) and divergence (fig. \ref{fig_divdacorr}.b) self-correlations $K_{\dot{a},\dot{a}}$ and $K_{\nabla,\nabla}$  together with the cross-correlation $K_{\dot{a},\nabla}$ (fig. \ref{fig_divdacorr}.c). In figure \ref{fig_divdacorr}.d the average of positive and negative $\tau$ of fig. \ref{fig_divdacorr}.c is made. 
A strong decorrelation at the first delay can be observed in $K_{\dot{a},\dot{a}}$. This probably is a consequence of the high level of noise present on the determination of $\dot{a}$, and not of a real lack of time correlation of the strain rate. White noise origin of the strong time decorrelation is confirmed by the fact that $\dot{a}$ and $\nabla\Vec{v}$ instead correlates in time (\ref{fig_divdacorr}.c), with a characteristic time comparable with the self correlation characteristic time $\tau_K$ of $\nabla\Vec{v}\left(t,\Vec{x}\right)$ (Fig. \ref{fig_divdacorr}.b).
The fact that an apparently delta-correlated signal - as the nuclear strain rate seems to be - correlates in time with the divergence might seem weird. As we show in the following, it is explained assuming that also nuclear strain rate has a finite time correlation hidden behind a strong delta correlated white noise.

For a generic quantity $q\left(t\right)$ whose real measurement $\tilde{q}\left(t\right)=q\left(t\right)+\sigma_q$  is affected by the delta correlated white noise $\sigma_q$, the non normalized self correlation is:

\begin{equation}
    \left<\tilde{p}\left(t\right) \cdot \tilde{p}\left(t+\tau\right)\right>_t = \left<p\left(t\right) \cdot p\left(t+\tau\right)\right>_t + \sigma_p^2  \delta \left(\tau\right)
\end{equation}

For the nuclear strain rate, it is possible to recover an indicative value of $\sigma_p$ from the $MSS$ as the square root of the offset $\sigma_w$. For the reported DCIS HC monolayer, $\sqrt{\sigma_w} = 0.07$. If we compare this value with the square root of the non normalized self-correlation evaluated in $\tau=0$, equal to $0.08$, we perceive how small is the hidden genuine correlation compared to white noise.  

If the large white noise impedes the observation of time correlation of nuclear strain rate, it does not hinder information on the correlation with independently measured variables like divergence. Indeed, in cross correlation between noisy data $\tilde{p}$ and $\tilde{q}$ with uncorrelated errors $\sigma_p$ and $\sigma_q$ it results:

\begin{equation}
    \left<\tilde{p}\left(t\right) \cdot \tilde{q}\left(t+\tau\right)\right>_t = \left<p\left(t\right) \cdot q\left(t+\tau\right)\right>_t
\end{equation}

as other terms vanish because of the delta correlation of noises $\sigma_p$ and $\sigma_q$.

Oscillations at negative values of correlation at finite $\tau$ in $K_{\dot{a},\dot{a}}$ and $K_{\nabla,\nabla}$, and consequently in $K_{\dot{a},\nabla}$, can be rationalized considering that, as \cite{kubo1966fluctuation}:

\begin{equation}
 \lim_{\tau\to\infty} \frac{\left<a\left(\tau|t\right)^2\right>_t}{2\tau} = \int_{0}^{\infty} \left<\dot{a}\left(t\right)\dot{a}\left(t+\tau\right)\right> \,d\tau  ,
\end{equation}

in order to reach a plateau of $MSS$ for $\tau\rightarrow\infty$ the integral of the time correlations must be zero and thus correlations must assume negative values.

From the cross-correlation, we recovered the characteristic time $\tau_k$ through an exponential fit excluding point in $\tau=0$. Inclusion of $\tau=0$ results in a failure of both exponential and stretched exponential fits. Apparently, a first very fast decorrelation arrived at time scales not accessible in our experiments. An example of the fit is superposed to data in figure \ref{fig_divdacorr}.d and the resulting decorrelation rates $1/\tau_k$ are reported in figure \ref{fig_divdacorr}.e as a function of time for the experiments on jamming monolayers.

\section{. Dynamics of flocking monolayers}
\label{appe_flocking}

\begin{figure}
    \centering
    \includegraphics[width=\columnwidth]{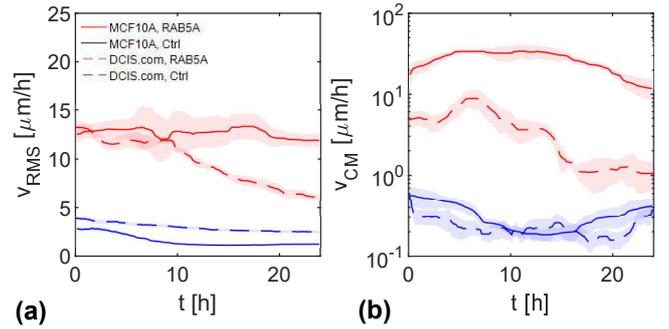}
    \caption{Time evolution of RMS velocity ${v}_{RMS}$ (\textbf{a.}) and of the modulus of the center of mass velocity (\textbf{b.}) obtained from PIV on fluorescent time lapses. Shadowed error bars are evaluated as the standard deviation of the mean evaluated for each sample over different FOVs. Reported data refers to flocking monolayers (blue for the control monolayers, red for the RAB5A overexpressing ones): MCF10A (continuous line) and MCF10A.DCIS.com (dashed line).}
    \label{fig_vrms_vcm_PIV_flocking}
\end{figure}

In figure \ref{fig_vrms_vcm_PIV_flocking} are reported the time evolution of $v_{RMS}$ (fig. \ref{fig_vrms_vcm_PIV_flocking}.a) and of the modulus of $\Vec{v}_{CM}$ (fig. \ref{fig_vrms_vcm_PIV_flocking}.b). As expected \cite{malinverno2017endocytic,giavazzi2018flocking}, both MCF10A and MCF10A.DCIS.com cell lines exhibit a reawakening of motility as a consequence of RAB5A overexpression, which is characterized by the emergence of a collective migration as a consequence of mutual velocity aligment of neighboring cells - testify by the motility peak in $\Vec{v}_{CM}$ - and by the acquisition of a fluid-like dynamical state in the center of mass reference system, as evident from the dramatic difference in $v_{RMS}$ between RAB5A overexpressing monolayers and control ones.



\bibliography{biblio_x}
\bibliographystyle{epj}
 
%

%
%

\end{document}